\documentclass{aa}  

\usepackage{graphicx}

\usepackage[desactivate]{aa_macros}
\usepackage{txfonts}
\usepackage[colorlinks=true, linkcolor=blue, urlcolor=blue, citecolor=blue]{hyperref}

\usepackage{natbib}
\usepackage{ulem}
\usepackage{xcolor}

\usepackage{soul}
\usepackage{longtable}
\usepackage{amsmath}
\usepackage{graphicx}
\def\code#1{\texttt{#1}}
\begin{document} 

   \title{Identifying the Mechanisms of Water Maser Variability During the Accretion Burst in NGC6334I}

   \author{J.M. Vorster\inst{1,2}
          \and
          J.O. Chibueze\inst{3,2,4}
          \and 
          T. Hirota\inst{5,6}
          \and
          G.C. MacLeod\inst{7,8}
          \and 
          D.J. van der Walt\inst{2}
          \and 
          E. I. Vorobyov\inst{9,10}
          \and 
          A.M. Sobolev\inst{9}
          \and 
          M. Juvela\inst{1}}

   \institute{Department of Physics, PO Box 64, FI-00014, University of Helsinki, Helsinki, Finland\\
              \email{jakobus.vorster@helsinki.fi}
         \and
             Centre for Space Research, Potchefstroom Campus, North-West University,
Potchefstroom 2531, South Africa
        \and
        Department of Mathematical Sciences, University of South Africa, Cnr Christian de Wet Rd and Pioneer Avenue, Florida Park, 1709, Roodepoort, South Africa
        \and
        Department of Physics and Astronomy, Faculty of Physical Sciences,  University of Nigeria, \\Carver Building, 1 University Road,
Nsukka 410001, Nigeria
        \and 
        Mizusawa VLBI Observatory, National Astronomical Observatory of
Japan, 2-12 Hoshiga-oka, Mizusawa, Oshu-shi, Iwate 023-0861, Japan
        \and 
        The Graduate University for Advanced Studies, SOKENDAI, 2-21-1 Osawa,
Mitaka, Tokyo 181-8588, Japan
        \and
        Department of Physical Sciences, The Open University of Tanzania, P.O. Box 23409, Dar-Es-Salaam, Tanzania
        \and 
        SARAO, Hartebeesthoek Radio Astronomy Observatory, PO Box 443, Krugersdorp, 1741, South Africa
        \and 
        Ural Federal University, 19 Mira Str., 620002 Ekaterinburg, Russia
        \and 
        Research Institute of Physics, Southern Federal University, Rostov-on-Don 344090, Russia
        }

   \date{Received September 15, 1996; accepted March 16, 1997}

 
  \abstract
   {High-mass young stellar objects gain most of their mass in short intense bursts of accretion. Maser emission is an invaluable tool in discovering and probing these accretion bursts.} 
   {We aim to observe the 22 GHz water maser response induced by the accretion burst in NGC6334I-MM1B and to identify the underlying maser variability mechanisms.}
   {We report seven epochs of Very Long Baseline Interferometry (VLBI) observations of 22 GHz water masers in NGC6334I with VLBI Exploration of Radio Astrometry (VERA) array, from 2014 to 2016, spanning the onset of the accretion burst in 2015.1. We also report 2019 Atacama Large Millimeter/submillimeter Array (ALMA) observations of 321 GHz water masers and 22 GHz single-dish maser monitoring by the Hartebeesthoek Radio Astronomical Observatory (HartRAO). We analyze long-term variability patterns and use proper motions with the 22 GHz to 321 GHz line ratio to distinguish between masers in non-dissociative C-shocks and dissociative J-shocks. We also calculated the burst-to-quiescent variance ratio of the single-dish time series.}
   {We detect a water maser distribution resembling a bipolar outflow morphology. The constant mean proper motion before and after the burst indicates that maser variability is due to excitation effects from variable radiation rather than jet ejecta. For the whole region, we find that the flux density variance ratio in the single-dish time series can identify maser efficiency variations in 22 GHz masers. The northern region, CM2-W2, is excited in C-shocks and showed long-term flaring with velocity-dependent excitation of new maser features after the onset of the burst. We propose that radiative heating of H$_2$ due to high-energy radiation from the accretion burst be the main mechanism for the flaring in CM2-W2. The southern regions are excited by J-shocks and showed short-term flaring and dampening of water masers. We attributed the diverse variability patterns in the southern regions to the radiative transfer of the burst energy in the complex source geometry.}  
%
   {Our results indicate that the effects of source geometry, shock type, and incident radiation spectrum are fundamental factors affecting 22 GHz maser variability. Investigation into water masers in irradiated shocks will improve their use as a diagnostic in time-variable radiation environments, such as accretion bursting sources.}
   \keywords{ISM: jets and outflows $-$ Stars: formation $-$ Masers $-$ Techniques: high angular resolution $-$ ISM: individual objects: NGC6334I}

   \maketitle
%

    \section{Introduction} 
\label{sec:introduction}
\subsection{Episodic Accretion in HMYSOs}
The mass-gain history of young stellar objects (YSOs) is characterized by variable accretion rates \citep{2014prpl.conf..387A,2023ASPC..534..355F}. Compared to steady-state accretion scenarios, variable accretion introduces complexities into our understanding of protostellar jets and envelope chemistry. For example, episodic accretion may be the cause of the clumpiness in high-velocity protostellar jets \citep{2014A&A...563A..87E,2018A&A...613A..18V,2019A&A...628A..68G}. Additionally, the periodic release of accretion energy by low-mass YSOs (LMYSOs, with M$_*<2 $ M$_\odot$) is known to affect the chemical abundances, ice and dust properties of the surrounding environment \citep{2009Natur.459..224A, 2016ApJ...821...46T,2022A&A...658A.191V}. It has been argued that protostars of all masses grow by disk accretion \citep{2016A&ARv..24....6B,2017NatPh..13..276C}. However, the mechanisms and environmental effects of episodic accretion by LMYSOs are much better understood than for their intermediate (M$_* = 2 - 8 $ M$_\odot$) and high-mass (M$_*>8 $ M$_\odot$) counterparts \citep{2021A&A...651L...3E,2023ASPC..534..355F,2023MNRAS.518..791E}. The larger mass gain of high-mass YSO (HMYSO) bursts may also lead to stronger effects on the chemistry of the surrounding environment than what is expected for LMYSOs \citep{2012ApJ...747...52D,2024A&A...684A..51G}. Hydrodynamic simulations have suggested that HMYSOs gain up to $50\%$ of their zero-age main-sequence mass in the bursting stage of accretion, which constitutes $<2\%$ of the stellar formation times \citep{2019MNRAS.482.5459M,2021MNRAS.500.4448M}. On the other hand, LMYSOs only gain $5\%-35\%$ of their final mass in bursts \citep{2012ApJ...747...52D}. To understand how episodic accretion in HMYSOs impacts the broader theory of star formation, we must investigate its rates, underlying mechanisms, and effects on the surrounding environment. There are theoretical predictions to the number and magnitude of accretion bursting events for different HMYSO masses, but the modelling is done without magnetic fields, and for elementary initial geometries \citep{2021MNRAS.500.4448M}.
Only about four accretion-bursting HMYSOs have been studied in-depth during the onset of their accretion bursts: S255IR-NIRS3, NGC6334I-MM1B, G358.93$-$0.03 and G24.33+0.14 \citep{2017ApJ...837L..29H,2017NatPh..13..276C,2020NatAs...4..506B,2022PASJ...74.1234H}. There are also a few HMYSOs with suspected accretion bursts, based on archival maser or infrared observations: V723 Car, G323.46-0.08 and M17 MIR \citep{2015MNRAS.446.4088T,2019MNRAS.487.2407P,2021ApJ...922...90C}. Disks around massive stars are difficult to observe, and it has been theoretically challenging to reproduce the rise times, accretion rates and accretion luminosities of the observed HMYSO bursts with mechanisms proposed for LMYSO bursts \citep{2021A&A...651L...3E}. Finally, the consequences of observed intermittent matter and energy ejections into the surrounding environment need to be investigated \citep{2018A&A...612A.103C,2020NatAs...4..506B}.
\subsection{The Unexpected Contemporaneous Flare of 6.7 GHz Methanol and 22 GHz Water Masers in NGC6334I}
\label{subsec:introduction_maser_variability}
All of the HMYSO bursts were detected by follow-up observations prompted by the flaring of the 6.7 GHz methanol maser transition \citep{2015ATel.8286....1F,2018MNRAS.478.1077M,2019ATel12446....1S}. A coordinated effort to detect and investigate bursting HMYSOs is being led by the Maser Monitoring Organization \citep[M2O,][]{2022evlb.confE..19B}. Single-dish monitoring of multiple maser lines has shown to be effective in identifying accretion bursts. The radiatively pumped 6.7 GHz methanol maser transition occurs in hot (T$_\text{dust} > 150$ K) and moderately dense ($10^5 < $ n$_\text{H$_2$} < 10^9$ cm$^{-3}$) gas \citep{2002MNRAS.331..521C,2005MNRAS.360..533C}. The correlation of 6.7 GHz methanol masers with the dust temperature implies that these masers could serve as an indicator and a "timer" tracking the evolution of the radiation for accretion bursts in HMYSOs \citep[e.g.][]{2018MNRAS.478.1077M}. It must be noted that a 6.7 GHz maser flare on its own is not enough to confirm an accretion burst and contemporaneous infrared and/or millimetre increase should also be detected \citep{2017NatPh..13..276C,2021ApJ...912L..17H,2021A&A...646A.161S}.
Another bright and common maser species that have been observed to flare during accretion bursts is the 22 GHz water maser transition \citep[e.g.][]{2018MNRAS.478.1077M,2018ApJ...866...87B,2022A&A...664A..44B}. In high-mass star-forming regions (HMSFRs), the 22 GHz water maser transition is typically understood to be collisionally pumped in dense ($10^8 < $ n$_\text{H$_2$} < 10^{10}$ cm$^{-3}$) shocked gas. The masers have different properties based on whether they originate in warm (T$_{\rm kin} \sim 400$ K) post-shock gas from a high-velocity (v$_{\rm shock} > 40$ km s$^{-1}$) dissociative J-shock or in hot (T$_{\rm kin} \sim 1000$ K) gas within a low velocity (v$_{\rm shock} < 40$ km s$^{-1}$) non-dissociative C-shock \citep{1992ApJ...394..221E,1996ApJ...456..250K,2013ApJ...773...70H}.

The most detailed models of water masers in J-shocks and C-shocks, as well as simpler broad parameter searches, all neglect the effects of external ultraviolet radiation \citep{1996ApJ...456..250K,2013ApJ...773...70H,2016MNRAS.456..374G}. Preliminary theoretical calculations of water masers in irradiated gas have suggested that an increase in stellar radiation dampens the effectiveness of the 22 GHz water maser pump \citep{2013JPhCS.461a2009N}. Observationally, an inverse correlation between 6.7 GHz methanol masers and 22 GHz water masers have been found in G107.298+5.639, G025.65+1.05 and the masers $<$ 1000 au from the bursting source NGC6334I-MM1B \citep{2016MNRAS.459L..56S,2018ApJ...866...87B,2019RAA....19...38S}. On the other hand, it was also observed that in NGC6334I some velocity components of  22 GHz water and 6.7 GHz methanol masers flared within 0.2 yr of each other \citep[Figure 9 of][]{2018MNRAS.478.1077M}. A lag of two months for the onset of 6.7 GHz methanol and 22 GHz water maser flares can be significant, as in G358.93-0.03 MM1 \citep{2022AJ....163...83B}. However, in the case of NGC6334I, large random variations in the 22 GHz water maser time series do not allow a determination in the flare onset with two-month precision. The contemporaneously flaring masers were at a similar velocity, but interferometric observations have revealed that they are likely not co-spatial. The 22 GHz water masers and the 6.7 GHz methanol masers are around 2750 au and 2000 au north of the bursting source MM1B respectively \citep{2018ApJ...854..170H,2018ApJ...866...87B}. Nonetheless, the precise mechanisms leading to the contemporaneous flaring of two maser species in NGC6334I that have contrary pumping mechanisms have not yet been adequately addressed. Previous studies have proposed that the flaring of the 22 GHz water masers in sources like NGC6334I, S255IR-NIRS3, and G358.93$-$0.03 was caused by amplified radiation fields \citep{2018MNRAS.478.1077M,2021A&A...647A..23H,2022A&A...664A..44B}. This explanation has invoked both collisional pumping as well as radiative pumping of 22 GHz water masers in conditions with T $_{\rm dust}> 1400$ K, T$_{\rm kin}< 500$ K and a narrower density range than collisionally pumped water masers \citep{2018ApJ...866...87B,2022MNRAS.513.1354G}.
\subsection{Millimetre Water Masers}
Apart from the 22 GHz transition, water also has many other maser lines such as 183 GHz, 321 GHz and 325 GHz. The most up-to-date list of water maser detections can be found at \url{www.maserdb.net} \citep{2022AJ....163..124L}. For predictions of maser lines, see \citet{2016MNRAS.456..374G}. Simultaneous observations of multiple co-spatial maser lines, combined with model predictions, can allow us to constrain hydrogen densities, kinetic temperatures, dust temperatures, and water abundances. The $10_{\rm 2, 9} \rightarrow 9_{\rm 3, 6}$ 321.225677 GHz (hereafter 321 GHz) water maser line can be combined with the 22 GHz water maser to distinguish between different types of shocks. It is rare compared to the 22 GHz transition, but has been detected with single-dish (e.g. W3(OH), W49N and W51 IRS 2) and interferometric observations \citep[e.g. Cepheus A and Orion I,][]{1990ApJ...350L..41M,2007ApJ...658L..55P,2014ApJ...782L..28H}. The 321 GHz water line has an upper energy level at $E_u/k = 1861$ K, which implies that it forms at high kinetic temperatures ($T \sim 1000$ K). Suppose that at 321 GHz and 22 GHz, the water masers in a particular source are co-spatial, then emissivity ratio $R = L_{21}/L_{321}$ can be used to discriminate between excitation from different shock types under the assumption of equal beaming for the two lines. For $R \sim 1$, it implies that the water maser is probably excited in C-shocks, where T$_{\rm kin}\sim 1000$ K \citep{1996ApJ...456..250K}. Otherwise, for $R \sim 10$ it implies J-shocks, where the masers form in T$_{\rm kin} =$ 400 K gas \citep{2013ApJ...773...70H}. The line ratios can also be combined with maser models to constrain other maser physical conditions \citep[e.g. ][]{1997MNRAS.285..303Y,2016MNRAS.456..374G}. The first detection of 321 GHz masers in an accretion-bursting source was in S255IR-NIRS 3 by \citet{2021A&A...647A..23H}. They found regions with both high $R \sim 10$ and low $R \sim 1$. This shows a variety of shock types which excite water masers in star-forming regions, which might respond differently in the time-variable radiation environments of accretion-bursting sources. 
\begin{figure*}
    \centering
    \includegraphics[width=0.9\textwidth]{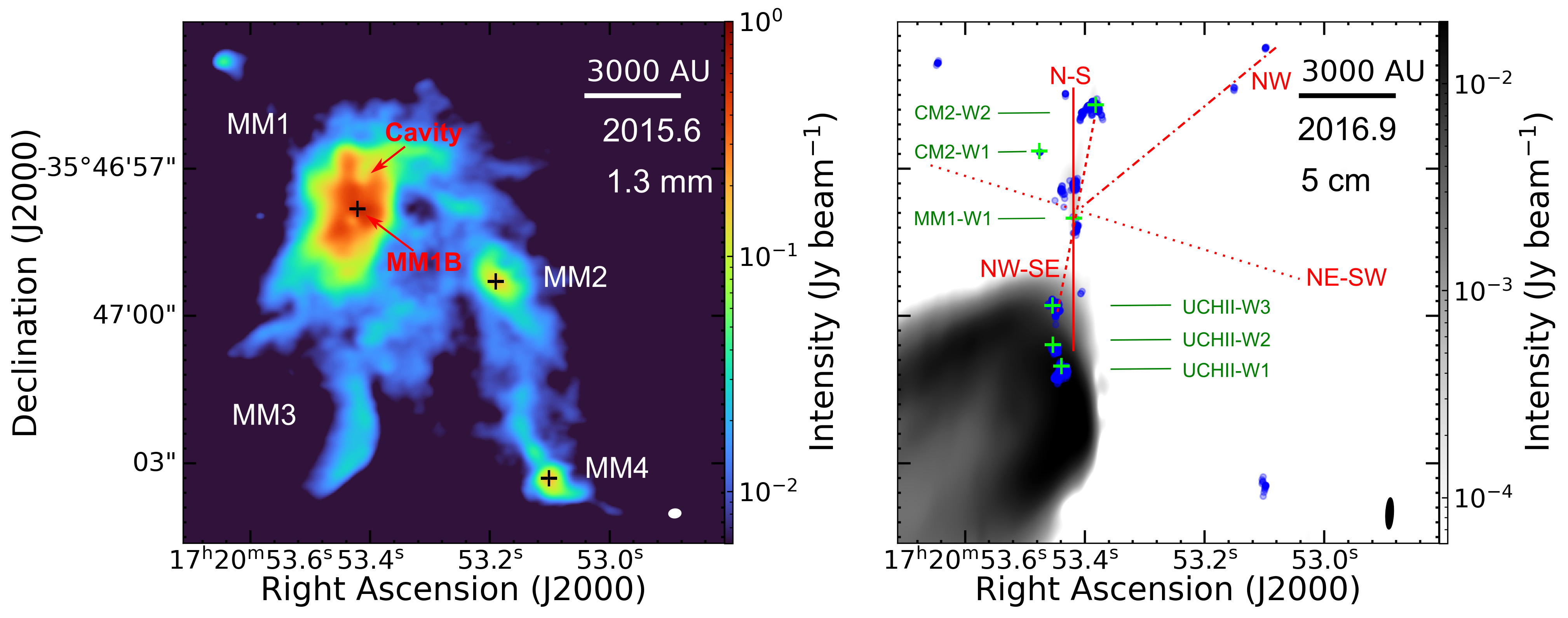}
    \caption{Overview of the source. Left: ALMA 2015.6 1.3 mm dust continuum from \citet{2016ApJ...832..187B}. The main regions MM1$-4$ are labelled in white. The peak positions of the bursting source, MM1B as well as MM2 and MM4 are marked with crosses. The location of the millimetre cavity is marked in red. Right: JVLA 2015.9 5 cm continuum with 2017.8 22 GHz water masers overlaid in blue \citep{2018ApJ...866...87B}. In green are the positions and names of the water maser associations introduced in Section \ref{subsec:introduction_target_source}. The positions and orientations of the N$-$S, NW \citep[solid and dashed-dotted respectively, from][]{2018ApJ...866...87B} and NW-SE \citep[dashed line, from][]{2021ApJ...908..175C} jets as well as the 0.5 pc NE-SW outflow \citep[dotted line, from][]{2011ApJ...743L..25Q} are marked in red lines.}
    \label{fig:source_overview}
\end{figure*}
\subsection{Target Source: NGC6334I}
\label{subsec:introduction_target_source}
Our target, NGC6334I \citep[$d = 1.3$ kpc][]{2014ApJ...784..114C, 2014A&A...566A..17W} is a complex HMSFR. It contains nine millimetre sources MM1$-$9 \citep{2006ApJ...649..888H,2016ApJ...832..187B}. Figure \ref{fig:source_overview} shows an overview of the millimetre and centimetre continuum of the source. The largest star-forming clump, MM1, is a "hot multi-core", containing seven dust cores with T$_{\rm dust}> 100$ K and high column densities $> 1\times 10^{25}$ cm$^{-2}$ \citep{2016ApJ...832..187B}. The most compact core MM1B, with a radius smaller than 155 au, started undergoing an accretion burst in January 2015 \citep{2017ApJ...837L..29H,2018MNRAS.478.1077M}. Infrared observations with SOFIA combined with radiative transfer modelling have found that MM1B has a mass of 6.7 $M_\odot$ and is undergoing a burst with an accretion rate of $\dot{M} \geq 2.3\times 10^{-3} M_\odot$ yr$^{-1}$ that will last for $40 - 130$ yr \citep{2021ApJ...912L..17H}. 

NGC6334I-MM1 harbours a large-scale wide-angle outflow, and at least three jets \citep{2011ApJ...743L..25Q,2018ApJ...866...87B,2021ApJ...908..175C}. The right panel of Figure \ref{fig:source_overview} shows the positions, orientations and sizes of the NE-SW outflow and the N-S, NW-SE and MM1B-NW jets. The positions of 22 GHz water masers observed with the JVLA in 2017.8 and the names of the maser associations relative to the outflows and jets from \citet{2018ApJ...866...87B} are also shown. The NW-SE jet is associated with the CM2-W2 water maser association at the northern edge and UCHII-W3 at the southern edge. The NW-SE jet has a small inclination angle of $-6^\circ$ in the plane of the sky \citep{2021ApJ...908..175C}. On the other hand, the southern edge of the N-S jet is associated with UCHII-W1 and UCHII-W2 \citep[Figures 8 and 9 of][]{2018ApJ...866...87B}. The maser association MM1-W1 is very close to the bursting source MM1B and might be affected by multiple outflows \citep{2021ApJ...908..175C}. 
\subsection{Observational Aims}
We aimed to characterize the effect of the accretion burst in NGC6334I on 22 GHz water maser variability and sought to determine its underlying mechanisms.
We conducted VLBI monitoring observations of 22 GHz water masers in NGC6334I before, during and after the onset of the accretion burst, using the VLBI Exploration of Radio Astrometry array \citep[VERA,][]{2020PASJ...72...50V}.  Additionally, we report the first detection of 321 GHz water masers in NGC6334I with Atacama Large Millimeter/submillimeter Array (ALMA) Cycle 6 observations. Our observations are co-temporal with the burst onset, allowing us to track changes in water masers as the burst begins, in contrast to previous studies by \citet{2018ApJ...866...87B}. This work also extends the proper motion study of \citet{2021ApJ...908..175C} by covering a longer timeframe and considering a broader set of observations and theoretical background.
\section{Observations and data reduction} 
\label{sec:observations} 
\subsection{VERA Observations} 
\label{subsec:vera_observations}
\begin{table*}[ht]
    \centering
    \begin{tabular}{llllll}
    \hline
      Epoch  & Observation Date & Decimal Date & \multicolumn{2}{c}{Phase Centre} & $R_{V}$ \\
                    &                  &              & R.A. (J2000)        &  Decl. (J2000) & (km\,s$^{-1}$)  \\
      
      \hline
        1 & 19 Sep 2014 & 2014.72 & 17$^{\rm h}$13$^{\rm m}$09$^{\rm s}$.9415 & $-35^\circ$47$'$02$''$.200 & 0.42 \\
        2 & 25 Nov 2014 & 2014.90 & 17$^{\rm h}$13$^{\rm m}$09$^{\rm s}$.9415 &$-35^\circ$47$'$02$''$.200  & 0.42 \\
        3 & 31 Jan 2015 & 2015.08 & 17$^{\rm h}$20$^{\rm m}$53$^{\rm s}$.440 & $-35^\circ$47$'$02$''$.200  &0.42 \\
        4 & 14 Apr 2015 & 2015.28 & 17$^{\rm h}$20$^{\rm m}$53$^{\rm s}$.440 & $-35^\circ$47$'$02$''$.200 & 0.40 \\
        5 & 17 Nov 2015 & 2015.88 & 17$^{\rm h}$20$^{\rm m}$53$^{\rm s}$.378 & $-35^\circ$46$'$55$''$.847 & 0.42 \\
        6 & 09 Feb 2016 & 2016.11 & 17$^{\rm h}$20$^{\rm m}$53$^{\rm s}$.367 & $-35^\circ$46$'$56$''$.930 & 0.42 \\
        7 & 12 Mar 2016 & 2016.19 & 17$^{\rm h}$20$^{\rm m}$53$^{\rm s}$.367 & $-35^\circ$46$'$56$''$.930 & 0.42 \\
        \hline
    \end{tabular}
    \caption{Summary of VERA observations, containing the dates of each observation, the phase tracking centre and the spectral resolution $R_V$. The onset of the accretion burst was 2015.0 and the 6.7 GHz methanol maser peak was 2015.6 \citep{2018MNRAS.478.1077M}.}
    \label{tab:VERAObsSummary}
\end{table*}
We did seven epochs of observations of the 22.23508 GHz water maser transition with VERA between 2014 and 2016. These epochs cover the time before, during and after the onset of the accretion burst (see Table \ref{tab:VERAObsSummary}). NRAO530 was used as the phase and bandpass calibrator while J1713-3418 was used as the flux calibrator. The data were correlated with a software correlator in the NAOJ Mizusawa campus \citep{2016PASJ...68..105O}.
The spectral resolution was $0.42$ km\,s$^{-1}$ for all epochs and the beam size was $2.2\times 0.8$ mas with a position angle of $-13^\circ$. The data were reduced with standard data-reduction procedures with the Astronomical Image Processing System (AIPS) \citep{1996ASPC..101...37V}. For each epoch, self-calibration was applied to the bright $-7.6$ km\,s$^{-1}$ feature to improve the signal-to-noise ratio of the images.
After self-calibration, the line-free RMS noise, $\sigma_\text{RMS}$, was around 0.8 Jy\,beam$^{-1}$ for each epoch, but could be worse in channels with bright maser emission. Phase referencing was not successful in any observing session, due to sensitivity constraints from the low elevation of NGC6334I ($\delta = -34^\circ$) observed with VERA at a latitude $30^\circ - 40^\circ$. The loss of precision in absolute position does not affect any of our results. The absolute position on the sky was estimated to sufficient precision by finding the position of our $-7.6$ km\,s$^{-1}$ self-calibration reference feature in the more sensitive contemporaneous observations by \citet{2021ApJ...908..175C}. We took into account the differences in the velocity reference frame between the KaVA and VERA observations, as well as the proper motion of the feature (see Section \ref{subsubsec:proper_motion_calculations}). In Appendix \ref{subsec:discussion_pm} we compare our observations to those of \citet{2021ApJ...908..175C}.
\subsubsection{Maser Spots and Features}
\label{subsubsec:spots_and_features}
The maser maps were produced by fitting 2D Gaussian functions on the emission peaks in each channel with the \code{SAD} AIPS task. \code{SAD} erroneously identified imaging artefacts, such as sidelobe emission, as detections. Sidelobe emission was identified as faint maser spots that are spatially symmetric around a bright feature. Sidelobe spots also displayed non-Gaussian spectral profiles. For our observations, we found that grouping spots that extended over at least four contiguous channels and showed a Gaussian spectral profile showed a good balance of minimizing false detections and maximizing real detections. Maser spots are individual Gaussian fits to single channels, while maser features are physically grounded, have a full-width at half maximum of $0.5 - 2$ km\,s$^{-1}$ and are spread across multiple channels for observations which spectrally resolve the line. Maser features have a characteristic size of 1 au \citep[e.g.,][]{1994ApJ...429..253G,2001ApJ...560..853T,2008ApJ...685..285M}. The grouping of spots into features was done with the flux-weighted mean of maser spots, given by:
\begin{equation}
    \vec{x}_{\rm f} = \frac{\sum I_i \vec{x}_i}{\sum I_i}
    \label{eq:intensity_weighted_centroid}
\end{equation}
with $\vec{x}_i$ and $I_i$ the spot spatial coordinates and intensity  respectively. The uncertainties on the position of the feature were calculated with 10$^5$ Monte Carlo (MC) simulations for each feature. In each MC iteration, a position was sampled from a normal distribution with a width equal to the astrometric uncertainty of the maser spot, and Equation \ref{eq:intensity_weighted_centroid} was applied. The uncertainties reported in Table \ref{tab:maser_features} are 3$\sigma_{\rm RMS}$ of the $10^5$ MC runs for each feature. The uncertainties for maser features are $30 - 200$ $\mu$as in Right Ascension and $70 - 500$ $\mu$as in Declination. The SNR of the maser feature determined its positional uncertainty. The peak intensity, $I_{\rm peak}$, radial velocity $V_{\rm LSR}$ and full-width half-maximum (FWHM) of the maser feature was calculated with a Gaussian fit to the spectral profile of the maser feature. 
\subsubsection{Proper Motions}
\label{subsubsec:proper_motion_calculations}
We calculated relative proper motions for maser features that were persistent over consecutive epochs and with $V_{\rm LSR}$ within 0.5 km\,s$^{-1}$. Due to burst-induced variability, we also split the epochs into "pre-burst" (epochs 1$-$4) and "burst" (epochs 5$-$7). The accretion burst introduced variability in all epochs after 2015.0, but certain features were persistent until epoch 4 (2015.3). Relative proper motions were calculated with a linear fit to the position over time of the feature positions $\vec{x}_f$. The uncertainties were uncertainties on the fit.
 
Relative proper motions are not physical, as they are in the rest frame of the reference feature. Further, without absolute phase referencing, the proper motions cannot be put in absolute coordinates in the sky. We shifted the proper motions to an absolute position and velocity frame. 
 
The absolute position was obtained by shifting the reference frame to the reference feature in UCHII-W1 at ($\alpha$, $\delta$) $=$ ($17^{\rm h}20^{\rm m}52^{\rm s}.600$, $-35^\circ46'50''.508$) of 2015.95$-$2016.01 KaVA observations \citet{2021ApJ...908..175C}, that corresponds to the time between epoch five (2015.88) and six (2016.11) of our observations. This was done by setting the pre-burst and burst proper motions of our VERA reference feature equal to the proper motions of the same maser feature in the KaVA observations \citep[proper motion IDs: $70-73$ of Table 1 in][]{2021ApJ...908..175C}. The change in velocity frame was done by changing the spot maps, not the proper motions. It was done with the following equation:
\begin{equation}
    \vec{x}_{\rm shifted} = \vec{x}_{\rm unshifted} + \vec{\mu}_{\rm T} t_{\rm obs} + \vec{P}_{\rm T}
    \label{eq:frame_shifting}
\end{equation}
with $\vec{x}_{\rm (un)shifted}$ the (un)shifted position of the maser spots. The proper motion transformation vector $\vec{\mu}_{\rm T}$ in mas yr$^{-1}$ is the difference in the proper motion measurements of the same feature between our VERA observations and the contemporaneous KaVA observations. The time after the first epoch is $t_{\rm obs}$ in years. Lastly, $\vec{P}_{\rm T}$ is the change in offset to make the KaVA reference feature the offset zero-point. For the pre-burst proper motions $\vec{\mu}_{\rm T} =$ ($-1.9 \pm 0.3$, $17.3\pm0.4$) mas yr$^{-1}$ and for the burst proper motions $\vec{\mu}_{\rm T} =$ ($-2.0\pm 0.3$, $16.9\pm0.4$) mas yr$^{-1}$. The position shift was the same for the pre-burst and burst epochs with $\vec{P}_{\rm T} = $($-0.68921$, $5.28800$) arcseconds. The uncertainty in the proper motions due to the velocity frameshift changes in the following way:
\begin{equation}
    \Delta \vec{\mu} = \sqrt{(\Delta \vec{\mu}_{\rm V})^2 + (\Delta \vec{\mu}_{\rm K})^2}
    \label{eq:uncertainty_shift}
\end{equation}
with $\Delta \vec{\mu}$ the uncertainty after the frame shift, $\Delta \vec{\mu}_{\rm V}$ mas yr$^{-1}$ the uncertainty of each VERA proper motion, and $\Delta \vec{\mu}_{\rm K}$ the uncertainty of the same feature measured in \citet{2021ApJ...908..175C}. After this shift, the proper motions are absolute in position, but not yet in velocity.
 
The proper motions were shifted to an absolute velocity frame by assuming that all the maser features in CM2-W2 have a mean speed equal to the upper limit of $3.8$ mas yr$^{-1}$ set by the JVLA observations of \citet{2018ApJ...866...87B}. In our case, the absolute velocity frame could not be calculated by assuming both poles of a bipolar outflow to have the same velocity \citep[as in e.g. ][]{2016MNRAS.460..283B}. If we assume the masers are moving at a position angle of $-79.4^\circ$ \citep{2021ApJ...908..175C}, it implies the upper limits ($\mu_\alpha < -0.7$ mas yr$^{-1}$, $\mu_\delta < 3.735$ mas yr$^{-1}$). These correspond to a proper motion shift of ($0.7$, $-15.292$) mas yr$^{-1}$ for the pre-burst proper motions and ($0.7$, $-15.301$) mas yr$^{-1}$ for the burst proper motions. Our proper motions are upper limits for northward proper motions and lower limits for southward proper motions. None of our conclusions are affected by uncertainties included by the shift to an absolute position and velocity frame.
\subsection{ALMA Band 7 Observations}
We observed NGC6334I with ALMA in Cycle 6 (PI: T. Hirota, 2018.1.00024.S) in band 7. The observations were carried out on 1 September 2019 toward ($\alpha$, $\delta$) $=$ (17$^{\rm h}$20$^{\rm m}$53$^{\rm s}$.44, $-35^\circ$47$'$02$''$.20) in configuration C43$-$7 with 48 antennas. The amplitude and bandpass calibrator was J1924-2914, while the phase calibrator was J1717-3342. The on-source time was 81.2 minutes. The pre-imaging calibration was done by the East-Asian ALMA Regional Center (EA-ARC) calibration pipeline. We then carried out phase-only self-calibration and imaging with the Common Astronomy Software Applications \citep[CASA, ][]{2007ASPC..376..127M}. We only report the 321.2 GHz narrow-band spectral window relevant to the 321.225677 GHz water maser for this work. Line data were obtained by subtracting the continuum in the UV domain. The spectral resolution was 0.245 MHz corresponding to a velocity resolution of 0.23 km\,s$^{-1}$. We regridded the spectra to a velocity resolution of 0.25 km\,s$^{-1}$. The beam size was 0.14$\times$0.10 arcsec with a position angle of $-78$ degrees. The CASA task \code{tclean} produced the spectral cubes with a robust clean parameter of 0.5 was used with a cell size of 0.0125$'$. After self-calibration, the line-free RMS noise was 27 mJy beam$^{-1}$. 
\section{Results}  
\label{sec:results}
\begin{table*}
    \centering
    \begin{tabular}{lllllll}
 \hline
  ID$^{\rm a}$ & Association &  \multicolumn{2}{c}{Offset$^{\rm b}$}        &   \multicolumn{3}{c}{Gaussian Parameters$^{\rm c}$}        \\
  & & $\alpha$ &   $\delta$  & $I_{\rm peak}$ &           $V_{\rm LSR}$ &        FWHM \\
  & & (arcsec) & (arcsec) & (Jy beam$^{-1}$) & (km\,s$^{-1}$) & (km\,s$^{-1}$) \\
\hline
 1-1 &      MM1-W1 &  0.41137 (0.00003) & -2.40979 (0.00007) &  111.1 (1.02) &  -2.82 (0.003) &  1.3 (0.01) \\
 1-2 &      CM2-W1 &  0.61499 (0.00005) & -0.73227 (0.00012) &   22.3 (0.34) & -10.39 (0.007) &  1.0 (0.02) \\
 1-3 &      CM2-W1 &  0.33346 (0.00003) & -0.12073 (0.00007) &   54.3 (0.61) &  -8.50 (0.010) &  1.3 (0.02) \\
 1-4 &      CM2-W2 & -0.00899 (0.00018) & -0.01210 (0.00047) &   10.7 (0.86) &  -8.04 (0.044) &  1.0 (0.14) \\
 1-5 &      CM2-W2 &  0.00155 (0.00007) &  0.00118 (0.00018) &    4.5 (0.11) & -11.80 (0.020) &  1.3 (0.05) \\
 \multicolumn{7}{c}{ ... } \\
 \hline
    \end{tabular}
    \caption{Properties of detected 22 GHz water maser features for all epochs. Uncertainties are shown in parentheses.
    $^{\rm a}$Identification of maser feature in the form EP-XX, with EP the epoch number and XX the maser ID for that epoch. $^{\rm b}$Offset in terms of the reference $-7.60$ km\,s$^{-1}$ feature. $^{\rm c}$As described in Section \ref{subsubsec:spots_and_features}. Uncertainties are on the fit. This is a shortened version of the original table to show it's form and content. The full table is available in electronic format.}
    \label{tab:maser_features}
\end{table*}
\begin{table*}
\centering
\begin{tabular}{llllllll}    
\hline
ID &    Region &        \multicolumn{2}{c}{Offset$^{\rm a}$} &        \multicolumn{2}{c}{Proper Motion$^{\rm b}$} &    $V_{\rm LSR}$ &                  Detections$^{\rm c}$ \\
 &  & $\alpha$  & $\delta$  & $\mu_\alpha$  & $\mu_\delta$  &   & \\
 &  & (arcsec)  & (arcsec)  & (mas yr$^{-1}$)  & (mas yr$^{-1}$)  & (km\,s$^{-1}$)  &  \\
 \hline
 1 &    CM2-W2 &  -0.35653  &   5.17924 &   2.6 (0.4) &    2.9 (1.1) &   -8.55 &            1100 [ 1-3 2-15] \\
 2 &    CM2-W1 &   -0.07502 &   4.56777 &   2.7 (0.5) &    2.5 (1.5) &  -10.38 &            1100 [ 1-2 2-14] \\
 3 &    MM1-W1 &   -0.27873 &   2.89027 &  -0.4 (0.1) &    2.4 (0.5) &   -2.77 &          1101 [1-1 2-2 4-2] \\
 4 &    MM1-W1 &   -0.27810 &   2.88906 &  -0.4 (0.2) &   -1.1 (0.7) &   -3.01 &              0011 [3-2 4-3] \\
 5 &    CM2-W2 &   -0.69000 &   5.29997 &  -1.3 (0.0) &    1.7 (0.1) &   -7.34 &  1111 [ 1-6 2-18 3-25 4-20] \\
 6 &    CM2-W2 &   -0.68813 &   5.30081 &  -0.7 (0.2) &   -1.8 (0.8) &  -11.14 &            0011 [3-27 4-21] \\
 7 &    CM2-W2 &   -0.66408 &   5.29855 &  -4.0 (0.3) &    1.2 (0.9) &  -20.08 &            0011 [3-21 4-25] \\
 8 &    CM2-W2 &   -0.66616 &   5.30894 &  -3.0 (0.3) &   14.7 (0.9) &  -25.45 &            0011 [3-20 4-27] \\
 9 &    CM2-W2 &   -0.64110 &   5.35781 &  -2.7 (0.2) &    5.3 (0.6) &  -14.75 &            0011 [3-31 4-30] \\
10 &    CM2-W2 &   -0.68843 &   5.30120 &  -0.9 (0.1) &    0.1 (0.5) &  -12.85 &  1111 [ 1-5 2-16 3-28 4-22] \\
11 &    CM2-W2 &   -0.68714 &   5.30214 &  -1.5 (0.3) &    2.4 (1.0) &  -13.14 &            0011 [3-26 4-23] \\
12 &  UCHII-W3 &    0.09454 &   1.30863 &   1.7 (0.3) &  -23.3 (1.0) &  -31.79 &              0011 [3-5 4-4] \\
13 &  UCHII-W1 &   -0.07786 &  -0.05531 &   0.2 (0.4) &   -2.2 (1.2) &  -15.21 &            1100 [1-13 2-11] \\
14 &  UCHII-W1 &   -0.07698 &  -0.05872 &  -1.2 (0.2) &    0.9 (0.7) &  -10.30 &            1001 [1-15 4-11] \\
15 &  UCHII-W1 &   -0.01769 &   0.04127 &   2.2 (0.1) &  -13.4 (0.5) &  -23.59 &        1110 [1-12 2-7 3-17] \\
16 &  UCHII-W1 &   -0.00633 &   0.01766 &   2.9 (0.1) &  -15.9 (0.3) &  -17.47 &        1110 [1-19 2-8 3-15] \\
17 &  UCHII-W1 &    0.10087 &  -0.16673 &  -1.0 (0.3) &  -13.4 (0.8) &  -21.63 &        1101 [1-17 2-22 4-9] \\
18 &  UCHII-W1 &    0.16962 &  -0.12107 &   4.2 (0.6) &  -14.5 (1.9) &  -34.04 &            1100 [1-16 2-23] \\
19 &  UCHII-W2 &    0.20347 &   0.46435 &   0.3 (0.5) &  -12.1 (1.6) &  -30.51 &             1100 [1-10 2-5] \\
20 &  UCHII-W2 &    0.20304 &   0.46359 &   0.0 (0.2) &   -8.4 (0.6) &  -30.64 &         1110 [1-11 2-6 3-7] \\
 \hline
 \end{tabular}
    \caption{Properties of the identified pre-burst proper motions. Uncertainties are in parentheses. $^{\rm a}$Offset in terms of the reference maser of \citet{2021ApJ...908..175C} at ($\alpha$,$\delta$) = ($17^{\rm h}20^{\rm m}52^{\rm s}.600$, $-35^\circ45'50''.508$). $^{\rm b}$In terms of the velocity frame described in Section \ref{subsubsec:proper_motion_calculations}. $^{\rm c}$Epochs of maser detections. The IDs in square brackets are the maser features from Table \ref{tab:maser_features} that were used in the fit.}
    \label{tab:pm_preburst}
\end{table*}
\begin{table*}
\centering
\begin{tabular}{llllllll}
\hline
ID & Region & \multicolumn{2}{c}{Offset} &        \multicolumn{2}{c}{Proper Motion} &    $V_{\rm LSR}$ &                  Detections \\
 &  & $\alpha$  & $\delta$  & $\mu_\alpha$  & $\mu_\delta$  &   & \\
 &  & (arcsec)  & (arcsec)  & (mas yr$^{-1}$)  & (mas yr$^{-1}$)  & (km\,s$^{-1}$)  &  \\
 \hline
 1 &    MM1-W1 &  -0.27355  &   2.90128 &    1.0 (0.7) &   -7.2 (2.4) &   -3.24 &         011 [6-2 7-2] \\
 2 &    MM1-W1 &  -0.27414  &   2.90409 &  -0.0 (0.8)  &   -1.6 (2.6) &   -2.61 &         011 [6-1 7-1] \\
 3 &  UCHII-W3 &    0.13716 &   1.28321 &   4.4 (0.8)  &  -24.2 (2.4) &  -36.75 &         011 [6-3 7-3] \\
 4 &  UCHII-W3 &    0.17808 &   1.28082 &   4.2 (0.8)  &  -24.3 (2.5) &  -48.53 &         011 [6-4 7-4] \\
 5 &  UCHII-W2 &    0.20707 &   0.46699 &  -0.6 (0.4)  &  -11.5 (1.2) &  -28.33 &         011 [6-5 7-5] \\
 6 &  UCHII-W1 &   0.16472  &  -0.14416 &   1.7 (0.9)  &  -15.5 (2.8) &  -16.15 &        011 [6-10 7-8] \\
 7 &  UCHII-W1 &  -0.00069  &   0.01222 &   2.5 (0.3)  &  -18.3 (0.8) &  -21.07 &       111 [3 6-6 7-6] \\
 8 &    CM2-W2 &  -0.73733  &   5.22028 &  -0.4 (1.9)  &   -0.9 (5.9) &   -6.64 &        011 [6-11 7-9] \\
 9 &    CM2-W2 &  -0.60153  &   5.42343 &   -0.0 (0.7) &    5.0 (2.2) &  -39.98 &       011 [6-33 7-28] \\
10 &    CM2-W2 &  -0.69004  &   5.29996 &   -1.1 (0.1) &    1.6 (0.4) &   -7.60 &  111 [ 5-8 6-13 7-11] \\
11 &    CM2-W2 &  -0.62282  &   5.38883 &   -3.0 (0.3) &    2.0 (0.9) &  -17.57 &       110 [5-17 6-32] \\
12 &    CM2-W2 &  -0.66279  &   5.32902 &   -1.3 (0.3) &    3.3 (0.8) &  -24.23 &       011 [6-18 7-17] \\
13 &    CM2-W2 &  -0.63785  &   5.38017 &   -0.7 (0.5) &    5.1 (1.7) &  -14.66 &       011 [6-30 7-25] \\
14 &    CM2-W2 &  -0.62765  &   5.39141 &    0.4 (0.4) &    5.9 (1.3) &   -0.96 &       011 [6-31 7-27] \\
15 &    CM2-W2 &  -0.64374  &   5.37748 &    1.0 (0.6) &    9.4 (1.8) &  -15.68 &       011 [6-28 7-24] \\
16 &    CM2-W2 &  -0.64103  &   5.37551 &   -0.5 (0.3) &    4.6 (1.0) &  -12.50 &       011 [6-29 7-23] \\
17 &    CM2-W2 &  -0.68647  &   5.31583 &  -1.1 (0.7)  &   -0.1 (2.1) &  -10.57 &       011 [6-15 7-14] \\
18 &    CM2-W2 &   -0.66783 &   5.35867 &   -1.1 (0.9) &    3.5 (2.8) &   -3.99 &       011 [6-24 7-19] \\
19 &    CM2-W2 &   -0.64887 &   5.34824 &   -0.2 (0.4) &    8.8 (1.2) &  -18.80 &       011 [6-22 7-20] \\
20 &    CM2-W2 &   -0.66483 &   5.31181 &   -2.9 (0.4) &    2.6 (1.1) &  -24.72 &       101 [5-10 7-13] \\
21 &    CM2-W2 &   -0.66314 &   5.32870 &    0.8 (0.2) &   -3.5 (0.6) &  -24.75 &       101 [5-12 7-15] \\
22 &  UCHII-W1 &   -0.07247 &  -0.04942 &   -0.2 (1.6) &  -10.2 (5.1) &  -13.87 &         011 [6-9 7-7] \\
\hline
\end{tabular}
    \caption{As Table \ref{tab:pm_preburst}, but for the proper motions after the onset of the burst.}
    \label{tab:pm_burst}
    
\end{table*}
The overall distribution of water masers in the region was consistent with the bipolar structure reported by \citet{2018ApJ...866...87B}. We detected water masers in six distinct regions, previously identified and labelled as CM2-W1, CM2-W2, MM1-W1, UCHII-W1, UCHII-W2, and UCHII-W3\footnote{The nomenclature for the maser associations originates from \citet{2018ApJ...866...87B} and was also used by \citet{2021ApJ...908..175C}. The names of CM2-W2 and W1 were swapped between \citet{2018ApJ...866...87B} and \citet{2021ApJ...908..175C}. We follow \citet{2021ApJ...908..175C}.}. Figure \ref{fig:vera_spotmaps} shows the positions and radial velocities of these maser associations across the entire field. Detailed information about individual maser features, including their associations, positions, radial velocities, peak flux densities, and FWHM for epochs 1-7, is provided in Table \ref{tab:maser_features}.

According to \citet{2018MNRAS.478.1077M}, the onset of the accretion burst was 2015.01 and the methanol maser flare peak was 2015.62. Our observations allow us to see the evolution of the water maser emission before (2014.7 $-$ 2015.3) and after the onset of the burst (2015.9 $-$ 2016.2). Time dependence in spatial distribution and flux density was found in all the water masers, over scales from 6000 au (5$''$) down to 1 au (sub-milliarcsecond) scale. Figure \ref{fig:vera_subregions} shows zoom-ins on the positions and radial velocities of the water masers in the different associations over the seven epochs. Figure \ref{fig:velocities_per_association} shows the velocities over time for the maser features in each association. Figure \ref{fig:flux_density_time_series} shows the total line area (velocity integrated flux density) of each association over time. The bottom panel of Figure \ref{fig:flux_density_time_series} also shows their velocity range over time. The maser association CM2-W2 had many more maser features than the other associations.
\subsection{Maser variability in CM2-W2}
On a large scale, before the burst in 2014.7, CM2-W2 was only a few maser features around the systemic velocity of $-7.6$ km\,s$^{-1}$ spread over 100 au (Figures \ref{fig:vera_spotmaps} and \ref{fig:vera_subregions}). In the following epochs, the water masers progressively traced a bow shape with a size of $\sim$400 au. CM2-W2 also brightened over time, peaked at 2015.3, and stayed flaring for all epochs (Figure \ref{fig:flux_density_time_series}). 

There was also a velocity-dependent response in the maser features as the epochs progressed. As a whole, the velocity extent of the region increased linearly with time (Figures \ref{fig:velocities_per_association} and \ref{fig:flux_density_time_series}). Figure \ref{fig:cm2_velocity_spotmaps} shows the time dependence of maser features in different velocity bins over time. The size of the markers is proportional to the peak flux density of the maser features. The majority of the bright masers were detected at between $-15$ km\,s$^{-1}$ and 0 km\,s$^{-1}$. After the burst, these masers traced a linear structure spread over 0.27$''$ (356 au). The population of masers with $V_{\rm LSR}$ $\in [-30,-15]$ km\,s$^{-1}$ were only detected after the onset of the burst (2015.08) reaching a peak in intensity at 2015.88, forming a 0.16$''$ (210 au) linear structure with a clear NE-SW velocity gradient. In the epochs after 2015.88, these features reduced in brightness and linear extent, while retaining the velocity gradient. The $V_{\rm LSR}$ $\in [-30,-15]$ km\,s$^{-1}$ reached their peak one epoch (0.6 yr) later than the 6.7 GHz methanol masers and $-7.6$ km\,s$^{-1}$ water masers in CM2-W2. Lastly, two high-velocity features at $-40$ km\,s$^{-1}$ and $-50$ km\,s$^{-1}$ were only detected after 2015.9. Velocity-dependent time evolution was also seen in 2017.0, and 2017.8 \citep{2018ApJ...866...87B}. 

There were also variations in the water masers in CM2-W2 on scales of 1 au. Figure \ref{fig:reference_feature} shows the position and spectra of the reference maser feature. In 2014.7, the reference feature is a single-peaked feature with an extent of $460$ $\mu$as (0.6 au), which had been detected with JVLA in 2011 \citep{2018ApJ...866...87B}. The maser feature is unresolved, but the size of the maser feature is defined by the distance between the most red- and blue-shifted spots \citep[e.g.][]{1994ApJ...429..253G,2008ApJ...685..285M}. In 2014.9 and 2015.1, the feature was double peaked, while having the same extent on the sky. Between 2015.1 and 2015.3, the maser spots were over a larger extent of 1060 $\mu$as (1.4 au) and flared by a factor of four. The maser retained the large size and flaring state as the epochs continued. The spatial precision which allows us to detect the increase in the size of the masers is due to the high signal-to-noise (SNR $=$ 700) of the maser observations in the flaring state.
\subsection{Maser Variability in Other Associations}
\label{subsec:other_associations}
\begin{figure*}
    \centering
\includegraphics[width=\textwidth]{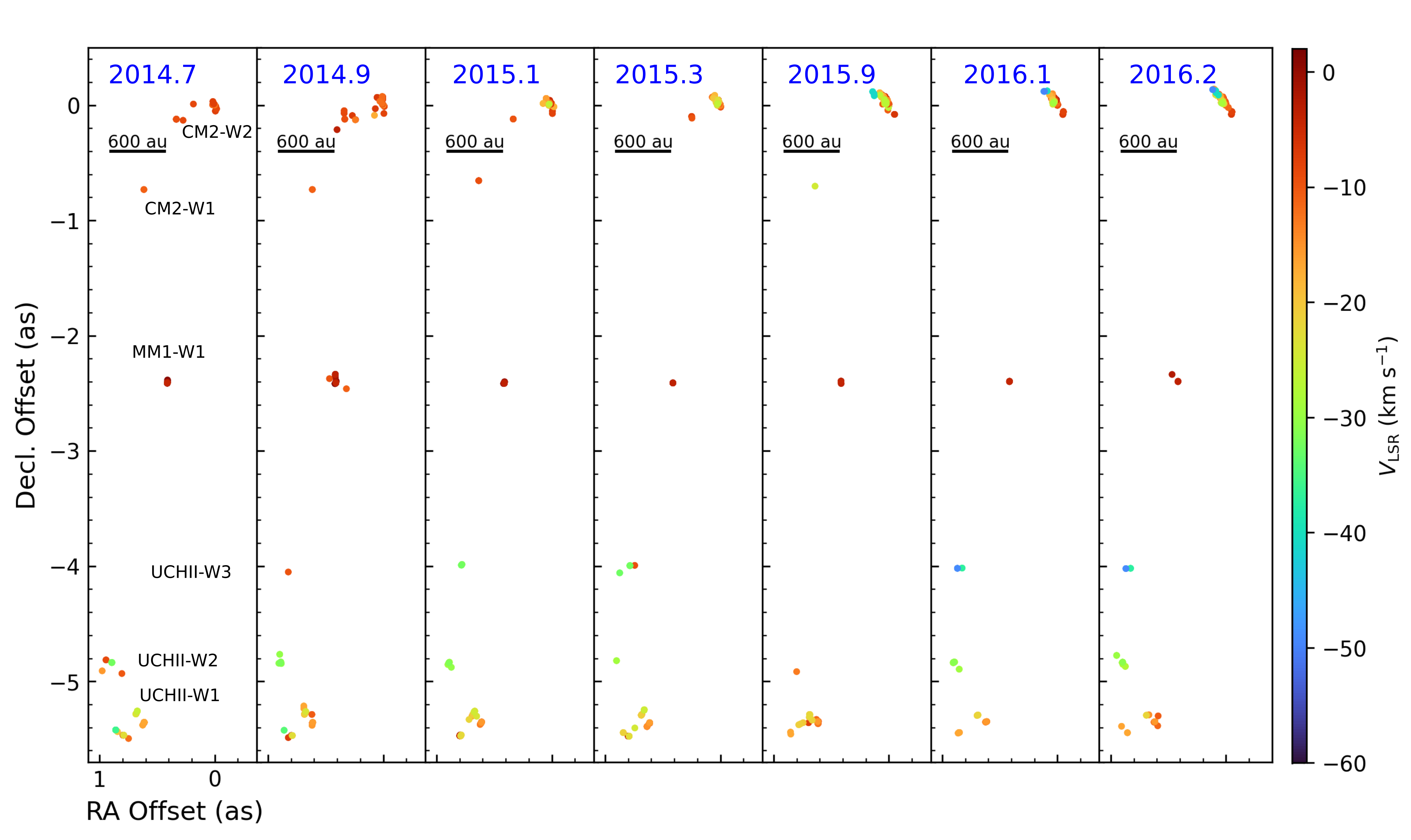}
    \caption{Positions and radial velocities of 22 GHz water masers at 2014.7 $-$ 2016.2 as observed with VERA. The names of the water maser associations are annotated in the leftmost panel. The linear scale for $d = 1.3$ kpc is shown in the top left of each panel. The accretion burst started at 2015.0 (after epoch 2), and peaked in 2015.6 (between epochs 4 and 5).}
\label{fig:vera_spotmaps}
\end{figure*}

\begin{figure*}
    \centering
    \includegraphics[width =\textwidth]{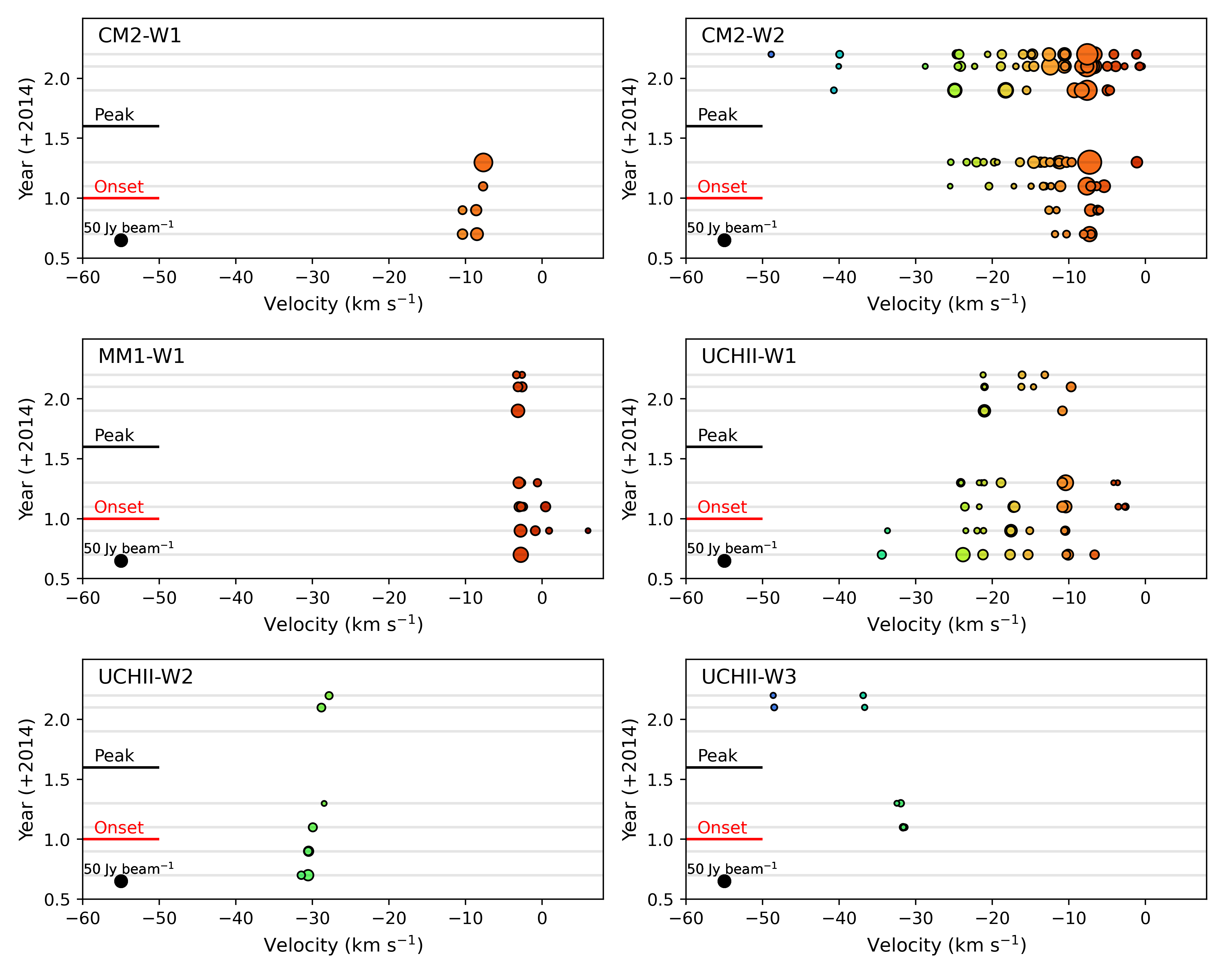}
    \caption{Centre velocities and peak intensities of maser features in each association over time. The marker size is proportional to the square of the feature peak intensity ($\propto I_{\rm peak}^{0.5}$). A scale for the marker size is shown in the bottom left of each panel. The markers are coloured with velocity. The grey horizontal lines indicate the dates of our observations. The dates of the onset and peak of the accretion burst traced by 6.7 GHz methanol masers is shown in red and black respectively \citep{2018MNRAS.478.1077M}.}
    \label{fig:velocities_per_association}
\end{figure*}

\begin{figure}
    \centering
    \includegraphics[width=0.45 \textwidth]{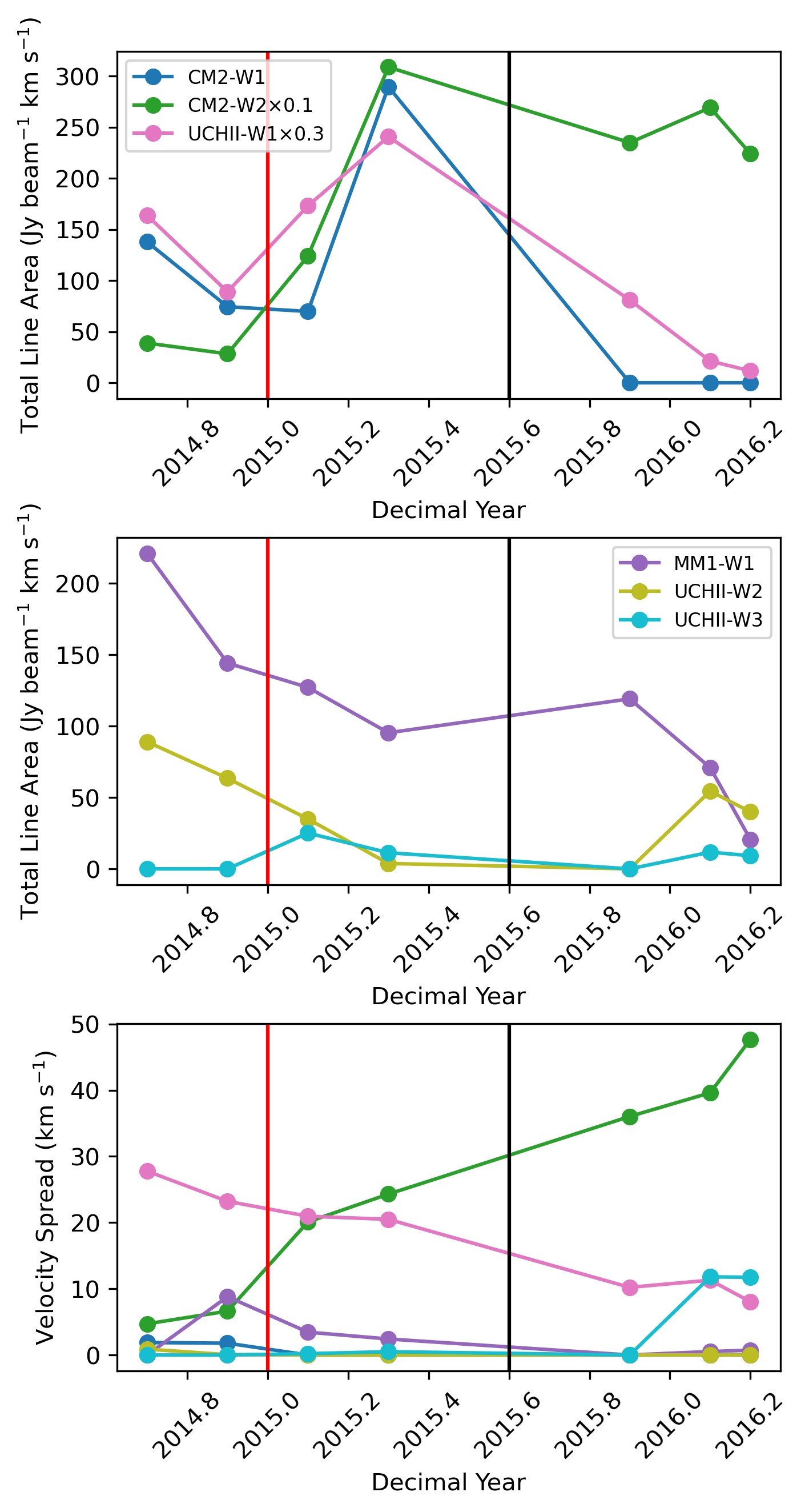}
    \caption{Time dependence of total velocity integrated intensity (line area) and velocity spread for each association from our VERA observations. The first and second panel shows line area for associations containing increasing and decreasing trends at the onset of the burst respectively. The third panel show the velocity spread for each association over time. The colours for the third panel are according to the legend in the first two panels. The red and black lines indicate the dates of the onset and peaks of the burst respectively.}
    \label{fig:flux_density_time_series}
\end{figure}

\begin{figure*}
    \centering
    \includegraphics[width=\textwidth]{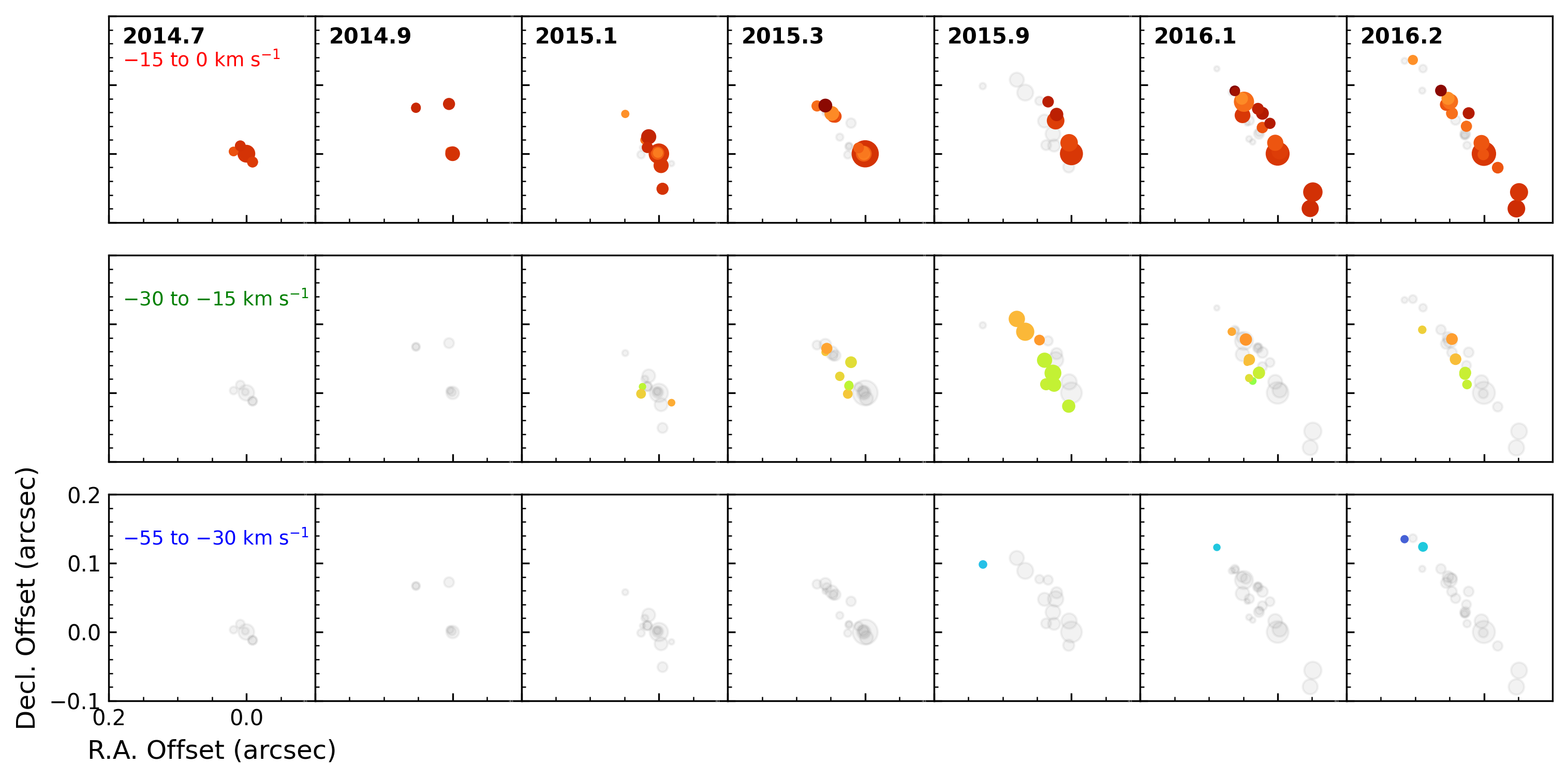}
    \caption{Velocity-dependent response of the water maser features in CM2-W2. The radius of the spots is proportional to the peak flux density. The rows indicate different velocity bins as indicated in the panels in the first column. Each column is the observation shown in the first row. Maser features which are outside of the respective velocity bin are showed in grey. The spatial scale for all panels is the same. The offset is in terms of the reference maser feature.}
\label{fig:cm2_velocity_spotmaps}
\end{figure*}

\begin{figure*}
    \centering    \includegraphics[width=\textwidth]{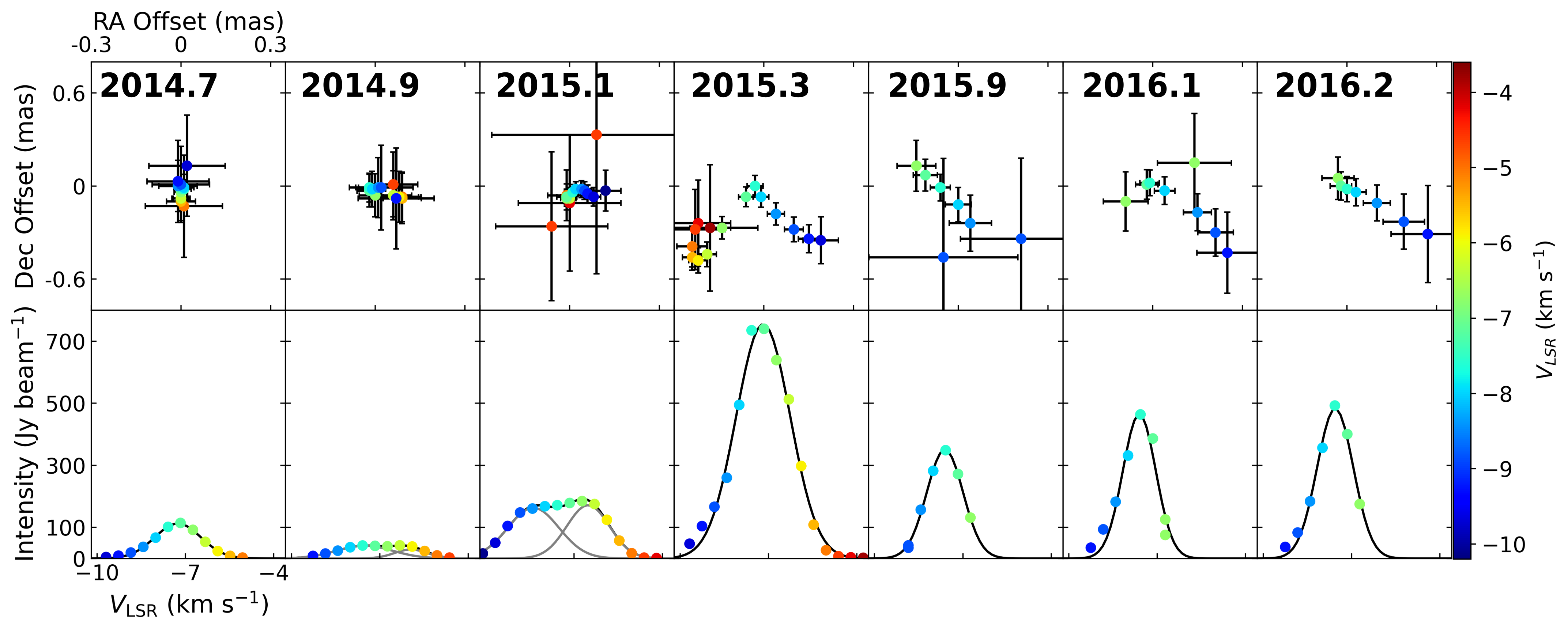}
    \caption{ Top panels: Offsets and $V_{\rm LSR}$ of maser spots in the brightest feature of CM2-W2 for each
epoch. The position, uncertainty and $V_{\rm LSR}$ of each spot is shown. Note the difference in spatial scale between Right Ascension and Declination. Bottom panels: Spectral profile of the maser spots in this feature, with Gaussian fits shown in black. In 2014.9 and 2015.1 the feature is well approximated by a double Gaussian.}
    \label{fig:reference_feature}
\end{figure*}

{\it CM2-W1} This association was detected in the first four epochs at $-9$ km\,s$^{-1}$ (2014-7 $-$ 2015.3, Figure \ref{fig:vera_spotmaps} and \ref{fig:velocities_per_association}). It peaked in flux density in 2015.3 (Figure \ref{fig:flux_density_time_series}), but was not detected after that.

{\it MM1-W1} This association is close to the bursting source. In epochs 2014.7 $-$ 2015.1 it formed a linear structure of $\sim 15$ au (Figure \ref{fig:vera_subregions}).  After 2015.3, the masers were displaced and then had a smaller linear scale of 7 au. The association monotonically decreased in line area and velocity range as the epochs progressed (middle panel of Figure \ref{fig:flux_density_time_series}).

\begin{figure*}
    \centering
    \includegraphics[width=\textwidth]{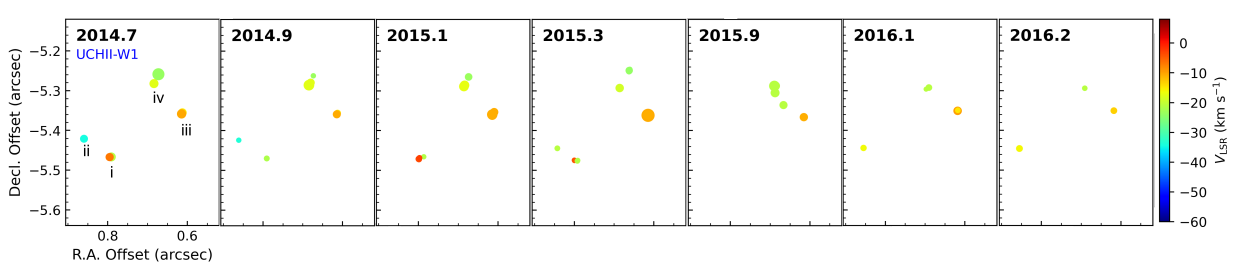}
    \caption{Velocity-dependent response of the water maser features in UCHII-W1. The area of the spots is proportional to the square of the peak flux density. The subregions of UCHII-W1, i$-$iv are marked in the first panel.}
    \label{fig:results_UCHII_W1}
\end{figure*}

{\it UCHII-W1} This association consisted of multiple maser associations spread over 1000 au at the southern point of the N-S molecular outflow detected by \citet{2018ApJ...866...87B}. As a whole, the region flared in flux density after the onset of the burst, peaking in 2015.3, after which it dramatically dropped in line area (Figure \ref{fig:flux_density_time_series}). The region also dramatically decreased in velocity spread from 30 km\,s$^{-1}$ in 2014.7 to 10 km\,s$^{-1}$ in 2016.2 (Figures \ref{fig:velocities_per_association} and \ref{fig:flux_density_time_series}). Figure \ref{fig:results_UCHII_W1} shows the four sub-associations, W1-i $-$ iv, numbered from south to north. The southernmost association, W1-i, with two features at $-$21.2 km\,s$^{-1}$ and $-$6.5 km\,s$^{-1}$ respectively, disappeared in 2015.3. The second sub-association W1-ii had a single maser feature at $-34.4$ km\,s$^{-1}$ which disappeared after the onset of the burst. A new maser feature at $-16.2$ km\,s$^{-1}$ was detected in 2016.1 and 2016.2. The third sub-association, W1-iii, was detected in all epochs around $-10$ km\,s$^{-1}$, but monotonically decreased in flux density as the epochs continued. The last sub-association, W1-iv, was also detected in all epochs, with monotonically decreasing flux densities. 

{\it UCHII-W2} This association consisted of a single maser feature at $-30$ km\,s$^{-1}$ before the burst. This feature monotonically decreased in line area from 2014.7 until disappearing in 2015.3 (middle panel of Figure \ref{fig:flux_density_time_series}). It was not detected in 2015.9, with a new double-peaked feature detected in 2016.1 $-$ 2016.2 at $-31$ km\,s$^{-1}$ and $-26$ km\,s$^{-1}$ displaced by $\sim 30$ au. The masers in UCHII-W2 were also slightly red-shifted in centre velocity as the epochs progressed (Figure \ref{fig:velocities_per_association}).

{\it UCHII-W3} This association was first detected after the onset of the burst at 2015.1 $-$ 2015.3 at $-31$ km\,s$^{-1}$, but it disappeared with two new features detected in 2016 at $-36$ km\,s$^{-1}$ and $-49$ km\,s$^{-1}$ separated by about 50 au. The features in this association showed a blue-shifting trend over time (Figure \ref{fig:velocities_per_association}).

\subsection{Pre-burst and Burst Proper Motions}
\label{subsec:results_pm}
\begin{figure*}
    \centering \includegraphics[width=0.8\textwidth]{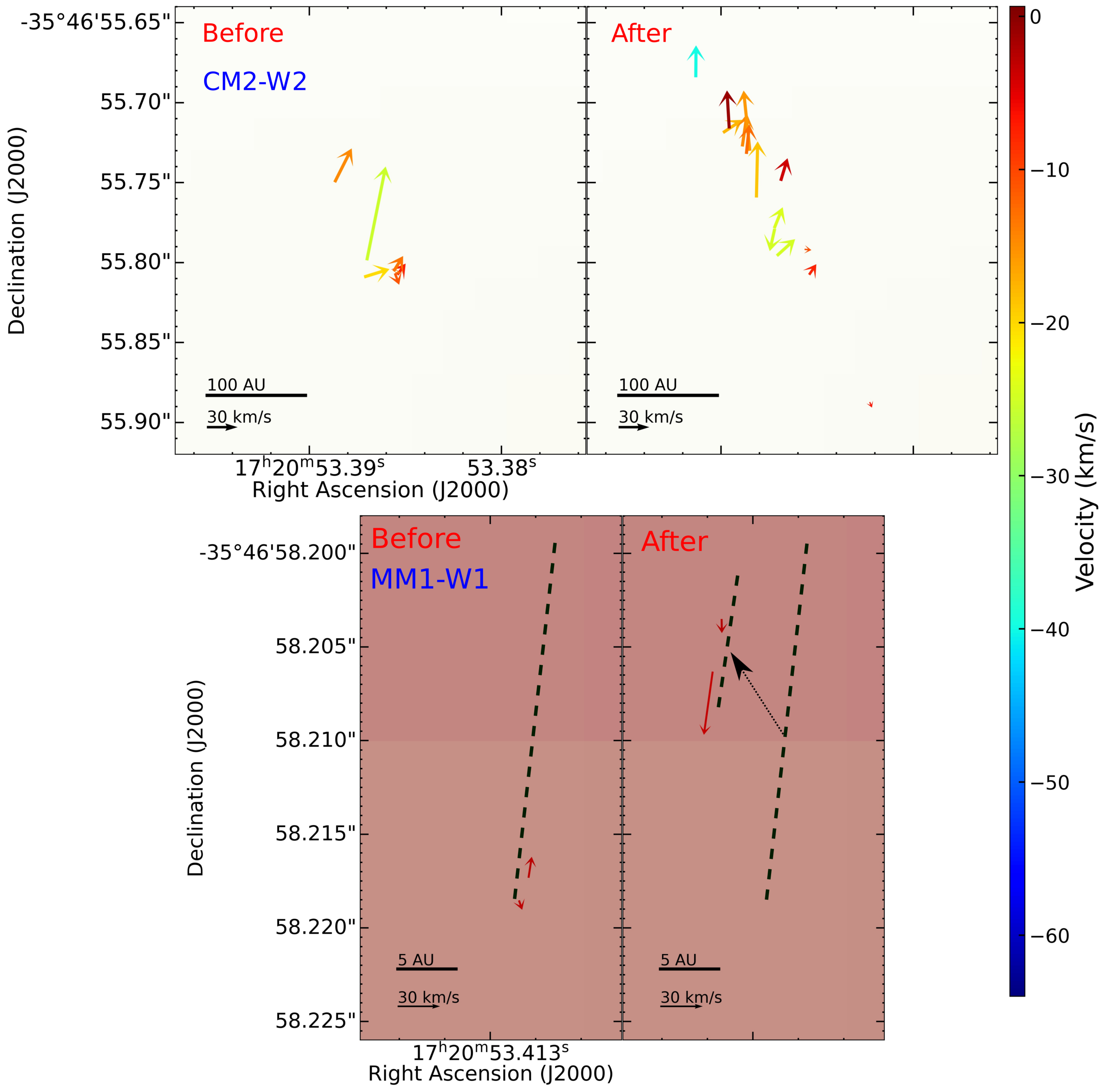}
    \caption{Proper motions of 22 GHz water masers in the northern regions of NGC6334I, before and
during the accretion burst. The positions,
orientations and colours of the arrows show the positions, proper motions and VLSR of the maser
features, respectively. The colour scale is on the right-hand side. The left panels show the pre-burst
epochs (2014.72 $-$ 2015.28) and the right panels show the burst epochs (2015.88 $-$
2016.19). The linear scales in position and space are shown in the bottom left corner of each panel.
In the bottom panel, black dashed lines show the position and orientation of the linear features shown
in the maser features, and the black arrow shows the displacement of the maser features after the
accretion burst.}
    \label{fig:proper_motions_north}
\end{figure*}

\begin{figure*}
    \centering
    \includegraphics[width=0.9\textwidth]{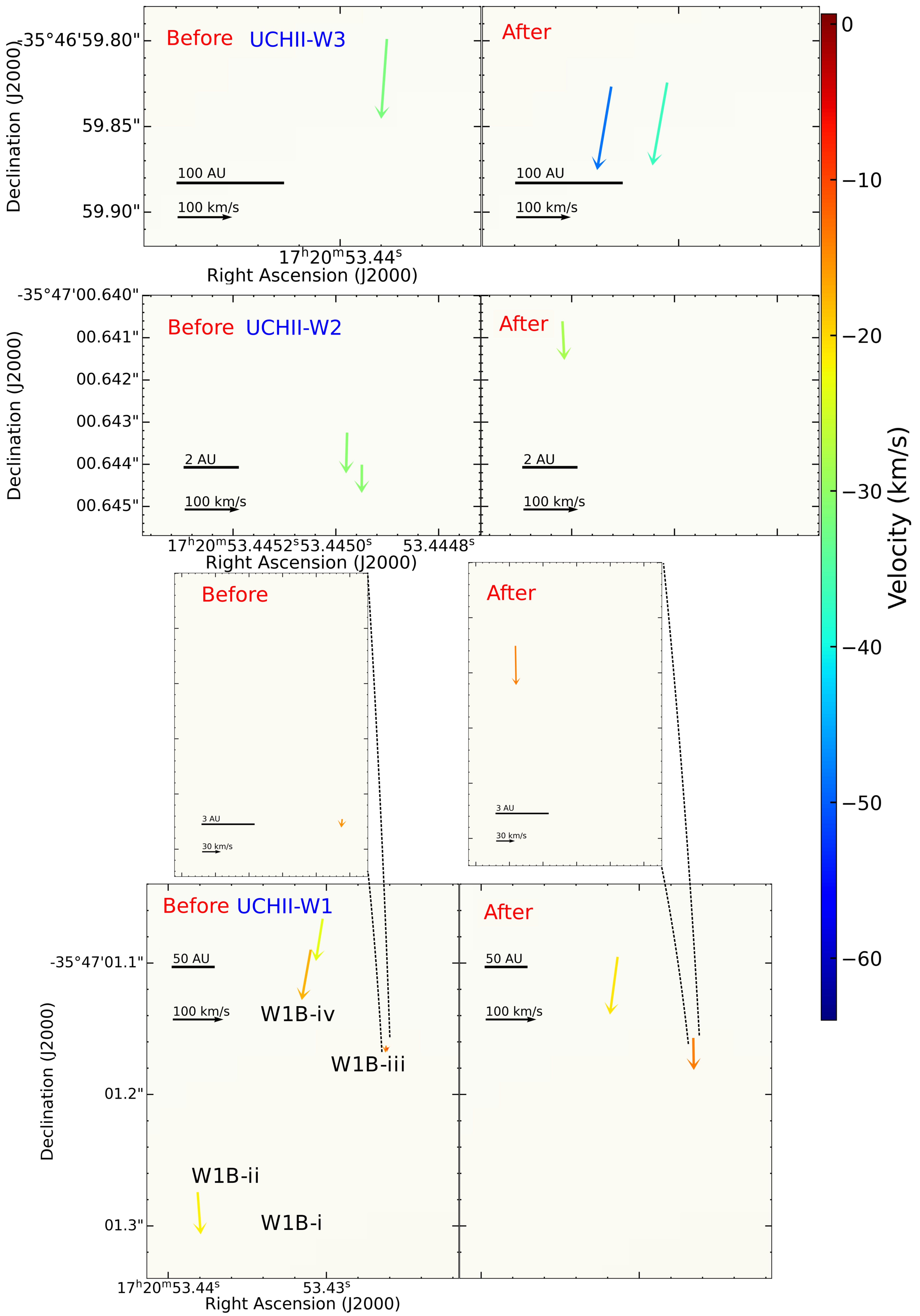}
    \caption{Like Figure \ref{fig:proper_motions_north}, but for the southern regions.}
    \label{fig:proper_motions_south}
\end{figure*}

We detected a total of 20 and 22 relative proper motions in the pre-burst (Table \ref{tab:pm_preburst}, $2014.72 - 2015.28$) and burst (Table \ref{tab:pm_burst}, $2015.88 - 2016.19$) epochs, respectively. Proper motions were detected in all associations where maser features were detected. Figures \ref{fig:proper_motions_north} and \ref{fig:proper_motions_south} show the proper motions for the northern regions (CM2-W2, MM1-W1) and southern regions (UCHII-W1, UCHII-W2 and UCHII-W3) respectively. There was only one proper motion in CM2-W1 that disappeared after the burst. The proper motions can be converted to transverse velocity with the relation $v_{\rm t} = 4.74\mu\cdot d$ with $v_{\rm t}$ in km\,s$^{-1}$, $\mu$ in mas yr$^{-1}$, $d$ in kpc and the factor of 4.74 arising from unit conversions. The mean speeds did not change with $\bar{v}_{\rm pre} = 50\pm40$ km\,s$^{-1}$ and $\bar{v}_{\rm burst} = 54\pm42$ km\,s$^{-1}$. After the burst, blue-shifted proper motions with $V_{\rm LSR}$ $\approx -50$ km\,s$^{-1}$ were detected. The mean speed of CM2-W2 was $33$ km\,s$^{-1}$ before and after the burst, with only the velocity dispersion increasing after the burst. This is expected as the reference feature was set to $30$ km\,s$^{-1}$ in all epochs to set the absolute velocity frame (see Section \ref{subsubsec:proper_motion_calculations}). The proper motions in CM2-W2 point in the NW direction, the same as the molecular outflow seen by \citet{2018ApJ...866...87B} and through velocity-variance covariance analysis of proper motions \citep{2021ApJ...908..175C}. In MM1-W1 the mean speeds increased from $v_{\rm MM1,pre} = 11\pm4$ km\,s$^{-1}$ to $v_{\rm MM1, burst} = 28\pm17$ km\,s$^{-1}$, and the centroid position of the proper motions shifted by 15 au. This is consistent with either maser dropping below detection limits and new masers being excited \citep[as seen in][ between 2017.0 and 2017.8]{2018ApJ...866...87B} or the same maser cloud undergoing a bulk displacement of 120 km\,s$^{-1}$ between 2015.3 and 2015.9. In the southern regions, almost all proper motions detected before the burst were not detected afterwards. In UCHII-W1, four associations named W1B-i$-$iv were detected. All of these associations had high speeds $v_{\rm pre} \sim 90$ km\,s$^{-1}$, except W1B-iii with $v = 17$ km\,s$^{-1}$. After the burst, only two high-velocity associations W1A-i (corresponding to W1B-iii) and W1A-ii (corresponding to W1B-iv) were detected. The association W1A-i was displaced by 12 au to the northeast, similar to MM1-W1. In UCHII-W2, there were southward pointing proper motions with $V_{\rm LSR}$ $=-30.5$ km\,s$^{-1}$ and $v_{\rm UCHII-W2, pre} = 70$ km\,s$^{-1}$ and a new feature at $V_{\rm LSR}$ $=-28.3$ km\,s$^{-1}$ with $v_{\rm UCHII-W2, burst} = 76$ km\,s$^{-1}$ displaced 6 au to the north-east. Finally, UCHII-W3 consisted of a single feature before the burst with $V_{\rm LSR}$$=-31.8$ and $v_{\rm UCHII-W3,pre} = 147$ km\,s$^{-1}$. After the burst, two new high-speed ($v = 160$ km\,s$^{-1}$) features at $V_{\rm LSR}$$= -48$ km\,s$^{-1}$ and $-36.8$ km\,s$^{-1}$ were detected, 61 au and 110 au displaced from the pre-burst feature.

In summary, the proper motions showed variability in all associations. High-velocity $> 70$ km\,s$^{-1}$ proper motions were detected in the southern regions, likely connected to either the N-S or NW-SE jets originating in MM1 \citep{2018ApJ...866...87B,2021ApJ...908..175C}. The speeds for the southern regions are lower limits, as the mean velocity of CM2-W2, $v = 30$km\,s$^{-1}$, is close to the upper limit of 23 km\,s$^{-1}$ from JVLA bulk motion between 2017.0 and 2017.8 \citep{2018ApJ...866...87B}.
\subsection{321 GHz Water Masers}
\begin{figure*}
    \centering
\includegraphics[width=0.9\textwidth]{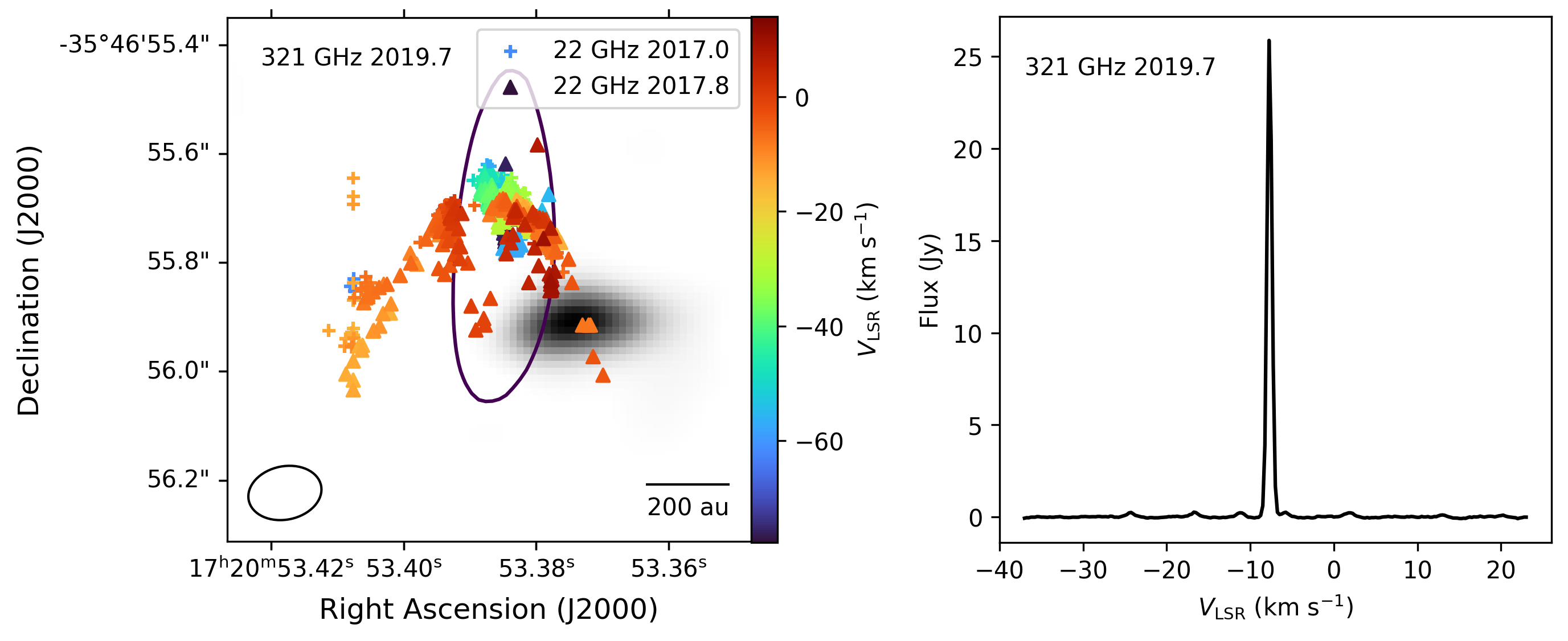}
    \caption{Left: Integrated intensity map of 2019.7 321 GHz water masers as detected with ALMA in greyscale. The coloured crosses and triangles shows the peak positions of 22 GHz water masers detected in 2017.0 and 2017.8 respectively while the purple contour is the 5 cm continuum point source \citep{2018ApJ...866...87B}. The ALMA beam and linear scale are shown in the bottom left and bottom right respectively. Right: Spatially integrated spectrum of the 321 GHz water masers. The small peaks are HCOOCH$_3$ thermal line emission, which is close to 321 GHz.}
    \label{fig:321GHz_masers}
\end{figure*}
\begin{table*}[ht]
    \centering
    \begin{tabular}{ccccc}
    Position$^{\rm a}$ & $V_{\rm LSR}$ & Integrated Flux & Peak Intensity & Deconvolved Size$^{\rm b}$ \\
     & (km s$^{-1}$) & (Jy) & (Jy beam$^{-1}$) & (mas $\times$ mas, deg) \\
\hline
    17$^{\rm h}$20$^{\rm m}$53$^{\rm s}$.3734, $-$35$^\circ$46$'$55$''$.9127 (1.1, 0.2) & $-8.25$ & 4.1 (0.05) & 2.5 (0.02)  & 157 $\times$ 41 (92) \\
    17$^{\rm h}$20$^{\rm m}$53$^{\rm s}$.3734, $-$35$^\circ$46$'$55$''$.9124 (0.6, 0.2) & $-8.00$ & 16.2 (0.11) & 9.8 (0.04)  & 156 $\times$ 41 (93) \\
    17$^{\rm h}$20$^{\rm m}$53$^{\rm s}$.3733, $-$35$^\circ$46$'$55$''$.9122 (0.6, 0.1) & $-7.75$ & 27.3 (0.17) & 16.5 (0.07)  & 156 $\times$ 39 (92) \\
    17$^{\rm h}$20$^{\rm m}$53$^{\rm s}$.3733, $-$35$^\circ$46$'$55$''$.9119 (0.6, 0.1) & $-7.50$ & 21.5 (0.13) & 13.0 (0.05)  & 155 $\times$ 41 (92) \\
    17$^{\rm h}$20$^{\rm m}$53$^{\rm s}$.3733, $-$35$^\circ$46$'$55$''$.9111 (0.6, 0.2) & $-7.25$ & 8.6 (0.06) & 5.2 (0.03) & 155 $\times$ 40 (92) \\
    \hline
    \end{tabular}
    \caption{Fits for 2019.7 321 GHz maser emission as observed with ALMA. $^{\rm a}$Positional uncertainty is in milliarcseconds. $^{\rm b}$Only upper limits for the size of an unresolved source.}
    \label{tab:millimetre_masers}
\end{table*}
We detected a single 321 GHz water maser feature at the southwestern end of CM2-W2. This is the first detection of 321 GHz water masers in NGC6334I. Figure \ref{fig:321GHz_masers} shows the integrated intensity of the 2019.7 321 GHz water maser, with the positions and velocities of 2017.0 and 2017.8 22 GHz water masers detected with JVLA \citep{2018ApJ...866...87B}. The 22 GHz water masers in CM2-W2 in 2011 were more localized than those observed in 2017.0, and not co-spatial with the 2019.7 321 GHz maser \citep{2018ApJ...866...87B}. The right panel of Figure \ref{fig:321GHz_masers} shows the spectrum of the maser, which peaks at $-7.75$ km\,s$^{-1}$ with an integrated flux of 27.3 Jy. The small peaks in the spectrum are thermal HCOOCH$_3$ lines close to 321 GHz \citep{2014PASJ...66..106H}. Table \ref{tab:millimetre_masers} shows the position, $V_{\rm LSR}$, integrated flux, peak intensity and deconvolved size from channel-by-channel 2D Gaussian fits to the maser feature. The maser feature is unresolved, showing the same elongation and size as the ALMA synthesized beam. Using the peak intensity and the upper limit for size reported in Table \ref{tab:millimetre_masers} the lower limit of the brightness temperature for the 321 GHz emission is 3.2$\times 10^{4}$ K, implying maser emission. The comparison with the closest published 22 GHz water maser observations with JVLA shows some maser spots from 2017.8 which are co-spatial and at the same velocity as the 2019.7 321 GHz water maser. It is interesting that in 2017.0 there were no water masers in the same position and frequency as the 321 GHz masers. As 22 GHz water masers are bright in a larger parameter space than 321 GHz water masers \citep{2016MNRAS.456..374G}, it could be that the 321 GHz water masers were only detectable between 2017.0 and 2019.7. We will consider the significance of the 321 GHz maser detection and the possible effect of time variability in CM2-W2 in Section \ref{subsec:water_maser_flare_cm2w2}.
\section{Discussion} 
\label{sec:discussions}
Our VLBI monitoring observations contemporaneous with the onset of the accretion burst give some clues to the mechanisms behind the water maser flare in NGC6334I. In Section \ref{subsec:discussion_interpreting}, we discuss the time variation of water masers in our VLBI observations taking the geometry of the various outflows and the neighbouring UCHII region into account. In Section \ref{subsec:variability_mechanisms} we use a toy model to distinguish three different sources of the variability and propose a statistical test on single-dish monitoring observations to distinguish mechanisms of water maser variability. Lastly, in Section \ref{subsec:water_maser_flare_cm2w2}, we aim to find the best explanation for the unexpected 22 GHz water maser flare in the bow-shock CM2-W2. In Appendix \ref{subsec:discussion_pm} we compare our proper motion calculations to previous estimates in NGC6334I and consider the conditions of reliability of proper motion measurements generally.
\subsection{Burst-Induced Water Maser Evolution} 
\label{subsec:discussion_interpreting}
\subsubsection{Source Radiation and Outflow Geometry of MM1}
\label{subsub:discussion_source_geometry}
Before we consider the effects of the accretion burst on the water masers, we will first consider the radiation spectrum from the accretion bursting source and the geometry of the complex outflows in MM1. The radiation spectrum for the protostar, disk, and cocoon environment of a bursting HMYSO is a complex open question. We can qualitatively consider the main radiation components of a protostar disk system to understand how episodic accretion drives diverse types of maser variability. An accretion-bursting source is expected to have infrared, optical, ultraviolet and possibly X-ray radiation components. Infrared radiation originates from the accretion disk and dusty natal cocoon \citep{2010ApJ...721..478H,2016A&ARv..24....6B}. Photospheric ultraviolet and optical radiation are expected to be bright, where low-optical depth cavities have been carved by jets and outflows, but faint when obscured by the disk which is often perpendicular to the jet axis \citep{2020AJ....160...78R}. During the burst, the photospheric SED reddens but becomes more luminous due to stellar bloating \citep{2019MNRAS.484.2482M}. It is also possible that bursting HMYSOs have an X-ray emission component. Some FU Orionis sources have much higher X-ray emission than quiescent T Tauri stars, but the effect is not universal \citep{2009ApJ...696..766S,2019ApJ...883..117K}. Further, intermediate to high-mass YSOs might have less X-ray emission, due to weak, complex magnetic fields \citep{2016ApJ...829...92A,2019A&A...622A..72V}. In summary, for most angles relative to the jet axis, infrared emission dominates the SED, but for some viewing angles, the high-energy component should not be neglected.

We consider the geometry of the natal envelope MM1 and its corresponding outflows to estimate which radiation components are incident on each maser association. To do this, we must first establish the relationship between MM3-UCHII, MM1, and the various outflows. This understanding is crucial because the large UCHII region can deflect outflowing material, but it is not immediately obvious whether MM3-UCHII and MM1 are at the same distance. Additionally, we consider the matter along the path from the bursting source to the maser association. High dust content in the intervening path will reprocess any high-energy radiation from the bursting source to infrared wavelengths. We can make qualitative inferences by considering the morphology of the outflows, jets and maser associations relative to MM3-UCHII. These are introduced in Section \ref{subsec:introduction_target_source} and Figure \ref{fig:source_overview}. Figures 8 and 9 of \citet{2018ApJ...866...87B} are also instructive. 

Firstly, MM3-UCHII must either be co-distant or behind MM1 for the free-free emission of the UCHII to provide seed photons for the maser associations UCHII-W1$-$3. We do see some influence of the UCHII region in deflecting the large-scale red-shifted lobe of NE-SW outflow \citep[Figure 9 of][]{2018ApJ...866...87B}. The undetected SE redshifted lobe of the NW-only jet might also be due to deflection from the UCHII region \citep{2018ApJ...866...87B}.
 
The N-S and NW-SE jets provide crucial information about the system's geometry. The NW-SE jet, with a small inclination angle of $-6^\circ$ \citep{2021ApJ...908..175C}, is associated with CM2-W2 and UCHII-W3 masers. Its SE lobe terminates at the bright edge of MM3-UCHII (Figure \ref{fig:source_overview}). The N-S jet's maser associations and thermal emission extend into MM3-UCHII's peak, suggesting MM3-UCHII might be behind UCHII-W1 and W2. This orientation could explain the lack of apparent deflection in the N-S jet. However, the effect of MM3-UCHII on the N-S jet remains tentative, with proper motion interpretations requiring careful consideration of velocity reference frames (see Appendix \ref{subsec:discussion_pm}).

Intervening dust and gas between MM1B and the maser associations is crucial for understanding the effects of accretion bursts on maser variability. A potential cavity SE of MM1B, visible in the 1.3 mm continuum (Figure \ref{fig:source_overview}), may allow a direct radiative connection between the bursting source and UCHII-W3. The N-S jet's cavities are not visible (Figure \ref{fig:source_overview}), but this might be the case if the N-S jet has a large inclination angle or if it has a different driving source than MM1B. This geometry could allow UCHII-W1 and W2 to receive only thermal and infrared components from the accretion burst due to dust reprocessing of high-energy radiation. However, these conclusions depend on the accuracy of our source geometry interpretations.

\subsubsection{Responses of Water Masers to the Burst in NGC6334I}
\label{subsub:cm2_w2_variability}
This section examines the response of water masers to an accretion burst in NGC6334I, focusing on several key maser associations. We will analyze the behaviour of CM2-W2, MM1-W1, UCHII-W1, UCHII-W2, and UCHII-W3, comparing their responses and exploring the underlying physical mechanisms. Table \ref{tab:summary_maser_behaviour} shows a summary of the observed trends in each maser association, predictions on the types of shocks exciting the water masers and estimates of each association's incident energy components.

The bow-shock CM2-W2 dominates the water maser emission in NGC6334I. Single-dish observations of the 22 GHz water masers with HartRAO by \citet{2018MNRAS.478.1077M} indicate that $>80\%$ of the post-burst flux density is between $-5$ km\,s$^{-1}$ and $-12$ km\,s$^{-1}$, which is dominated by CM2-W2 in our VLBI maps and the JVLA maps \citep{2018ApJ...866...87B}. The water maser features in CM2-W2 gradually increased in number, spatial extent and flux density from the onset of the burst (Figures \ref{fig:vera_spotmaps} and \ref{fig:cm2_velocity_spotmaps}). This process of gradually tracing more features has been observed up to two years after the burst \citep[2017.8,][]{2018ApJ...866...87B}. This indicates that whatever process is causing the maser flare in CM2-W2 at large ($> 50$ au) scales is continuous, starting at the onset of the burst, rather than a short ($< 1$ yr) single pulse "heatwave" as in G358.93-0.03 \citep{2020NatAs...4..506B,2022A&A...664A..44B}. Further, our VLBI observations confirm what was shown by \citet{2018ApJ...866...87B} that the large maser flares in the $-5$ km\,s$^{-1}$ to $-12$ km\,s$^{-1}$ velocity features seen by single-dish monitoring \citep{2018MNRAS.478.1077M} are the combination of multiple maser features spread spatially over $>$300 au. New persistent $-7$ km\,s$^{-1}$ features were excited south of CM2-W2 between 2015.88 and 2016.11 (top panel of Figure \ref{fig:cm2_velocity_spotmaps}). Different velocity features in CM2-W2 respond to the radiative changes due to the accretion burst at different times. The $-15$ km\,s$^{-1}$ to $0$ km\,s$^{-1}$ features (top panel of Figure \ref{fig:cm2_velocity_spotmaps}) respond immediately to the accretion burst, while the $-30$ km\,s$^{-1}$ to $-15$ km\,s$^{-1}$ features peak 0.6 yr later (middle panel of Figure \ref{fig:cm2_velocity_spotmaps}). The $< -30$ km\,s$^{-1}$ features flare even later, almost a year after the start of the burst (bottom panel of Figure \ref{fig:cm2_velocity_spotmaps}). Individual features show jumps in flux density and size, but overall there's a velocity-dependent continual "lighting up" of features (Figures \ref{fig:cm2_velocity_spotmaps} and \ref{fig:reference_feature}). This process continues at least 2.8 years after the accretion burst started \citep{2018ApJ...866...87B}. This behaviour could be explained by an irradiated surface layer which radiatively heats up immediately at the onset of the burst and then mixes with the rest of the gas at the shock front over multiple months.

The shock type in CM2-W2 can be inferred from observations of both 22 GHz and 321 GHz water masers. In the 2017.8 water masers, a maser spot was detected co-spatial and at the same velocity as the 321 GHz water maser reported in this work \citep[ID: 911 of Table 4 of ][]{2018ApJ...866...87B}. The 22 GHz spot has a reported integrated flux of 56 Jy, implying $R = F_{\rm 22}/F_{\rm 321} =  2.1$, which is more consistent with high temperature $T \sim 1000$ K maser emission from a C-shock \citep{1997MNRAS.285..303Y,2007ApJ...658L..55P}. This finding supports the interpretation of CM2-W2 as a C-shock region, which is crucial for understanding its unique response to the accretion burst. In Sections \ref{subsec:variability_mechanisms} and \ref{subsec:water_maser_flare_cm2w2}, we lay out a scenario in which variable radiation fields might explain the unique maser behaviour in CM2-W2.

The behaviour seen in CM2-W2 differs from that seen in other associations. In MM1-W1, \citet{2018ApJ...866...87B} identified a dampening of the water masers that was caused by the high dust temperature in the protocluster, which does not allow cold dust to be a sink and reduces the maser efficiency. Our observations of MM1-W1 during the accretion burst are consistent with the picture put forth by \citet{2018ApJ...866...87B}. We find that the fading of the water masers happened $<0.1$ yr after the onset of the burst (Figure \ref{fig:vera_spotmaps}). In UCHII-W2, a similar behaviour was observed.

 The behaviour of UCHII-W3 differed significantly from CM2-W2, despite both being radiatively connected to MM1B (Section \ref{subsub:discussion_source_geometry}). One explanation could be some unknown effects due to interactions with MM3-UCHII. Another could be the difference in shock types between UCHII-W3 and CM2-W2. UCHII-W3 likely features J-shocks (high-velocity proper motions of 150 km s$^{-1}$), while CM2-W2 shows characteristics of C-shocks (lower proper motions and $R = 2.1$). These two shock types have different chemical pathways through which water is produced and population inversion is sustained \citep{1996ApJ...456..250K,2013ApJ...773...70H}. The effects of UV irradiation on water maser emission for these shocks have not yet been explicitly modelled, but it has been argued that 22 GHz water masers from C-shocks are generally brighter and might have stronger responses to thermal and high-energy energy from the burst \citep{2013ApJ...773...70H,2024MNRAS.530.3342G}.

Why did CM2-W2 brighten and trace a larger velocity range after the burst, while UCHII-W1 brightened initially, but had reduced flux density and velocity range after that? The differences may lie in the shock type and source geometry. If there is no low-optical depth cavity, then UCHII-W1 only sees the thermal and infrared heat waves. The infrared heatwave is known to dampen masers rather than amplify them. On the other hand, UCHII-W1 also has high velocity ($> 70$ km s$^{-1}$) proper motions that indicate maser emission due to J-shocks. 

We can compare the water maser variability in NGC6334I to that of other sources. In S255IR-NIRS3, it was not clear that variability associated with the accretion burst was identified, as no 22 GHz maser monitoring was done before and after the accretion burst \citep{2021A&A...647A..23H}. The case of G358.93$-$0.03 is a noteworthy comparison with NGC6334I. In both cases, there was dampening of the masers close ($< 1500$ au) to the bursting source, as expected from maser theory as explained in Section \ref{subsec:introduction_maser_variability} \citep{2018ApJ...866...87B,2022A&A...664A..44B,2024AJ....167...63M}. Between the two sources, it is clear that the large scale ($> 10000$ au) density structure of the natal envelope modulates water maser variability. The density structure modulates the motion of the thermal heatwave, which is a pulse that can propagate at up to $\sim 4\%$ of the speed of light \citep{2020NatAs...4..506B}. For G358.93$-$0.03, water maser associations showed directionally dependent variability, which was attributed to inhomogeneous density distributions around the bursting source \citep{2024AJ....167...63M}.

Our analysis of NGC6334I reveals three fundamental considerations about water maser response to accretion bursts. The effect of thermal, infrared and high-energy components from accretion bursts affect water maser response separately. Then the intervening density distribution between the bursting source and the maser association modulates the burst energy components. Finally, water masers in C-shocks and J-shocks respond differently to these energetic components. In the next section, we consider different mechanisms of water maser variability. 
\begin{table*}[t]
    \centering
    \begin{tabular}{ccccccc}
        Association & Line Area Trend$^{\rm a}$ & Velocity Spread Trend & Shock Type$^{\rm b}$ & Incident Energy Components$^{\rm c}$ \\
         \hline
        CM2-W1  & flare then drop & monotonic decrease & J-type & thermal, infrared  \\
        CM2-W2  & flare then elevated & monotonic increase& C-type & thermal, infrared, high-energy \\
        MM1-W1  & monotonic decrease & flare then drop & C-type & thermal, infrared \\
        UCHII-W1 & flare then drop& monotonic decrease & J-type & thermal, infrared, high-energy? \\
        UCHII-W2 & monotonic decrease & monotonic decrease & J-type &  thermal, infrared, high-energy? \\
        UCHII-W3 & flare then drop &monotonic decrease & J-type & thermal, infrared, high-energy \\
        \hline        
    \end{tabular}
    \caption{Summary of the time-dependent behaviour of each maser association in NGC6334I between 2014.7 and 2016.1 compared to predictions of the shock types and incident energy components. $^{\rm a}$From Figure \ref{fig:flux_density_time_series}. $^{\rm b}$For CM2-W2 the estimate of the shock type is from considering the proper motions (Section \ref{subsec:results_pm}) and the 321 GHz to 22 GHz ratio (Section \ref{subsub:cm2_w2_variability}). For the other associations, the shock type prediction is only based on proper motions. $^{\rm c}$Based on the arguments from Section \ref{subsub:discussion_source_geometry}. Whether or not the high-energy components are incident on UCHII-W1 and W2 is dependent on whether there is a cavity between MM1B and these associations.}
    \label{tab:summary_maser_behaviour}
\end{table*}

\subsection{Variability Mechanisms in 22 GHz Water Masers}
\label{subsec:variability_mechanisms}
The inverse problem of calculating physical variables (H$_2$ density and temperature, H$_2$O abundance, shock velocity, maser path length, incident radiation) from 22 GHz water maser observations (position, radial velocity, flux density and time variability) is non-trivial. How do you know that your maser observations are informative about the underlying physical processes? For water maser observations to be informative, considering the observational degeneracy mentioned above, guidelines for judging the difference between fast "random" and slow "systematic" variability effects need to be considered.
\subsubsection{Three Sources of Variability}
To illustrate the three sources of variability in the collisionally pumped maser time series, we will adopt a simple toy model as an illustration. Consider a cylinder of velocity-coherent H$_2$O gas with a length $s$, a radius $a$, and a large aspect ratio of $s \gg a$. The gas has an optical depth $\tau = \kappa_{\nu}s$ at 22 GHz. The opacity, $\kappa_\nu$, is negative for masers, which amplify background radiation. The value of $\kappa_\nu$ for a molecule and volume element is dependent on the response of the level populations to the macroscopic variables of the gas. The maser saturates at a length $s_{\rm sat}$ and amplifies a background intensity $I_0$. For $s < s_{\rm sat}$, the maser amplification is exponential and for $s> s_{\rm sat}$ the amplification is linear. In the toy model, we take into account that $\kappa_\nu$ may behave differently in the saturated and unsaturated regimes. We also add random variations in the path length due to chaotic fluid flow in which collisionally pumped masers form, $\delta s = \epsilon(t) s$, where $|\epsilon(t)| < \epsilon_{\rm max}$ is a uniformly random value that changes with time. The distribution of $\epsilon_{\rm max} < 1$ is unknown. The observed intensity $I_\nu(s)$ for $s>s_{\rm sat} + \delta s(t)$ is then given by:
\begin{equation}
    I(s,t) = I_0(t)e^{\kappa_\nu,{\rm sat}(t) \cdot s_{\rm sat}}\Big[\kappa_\nu(t) \cdot\big(s - s_{\rm sat} + \delta s(t) \big) \Big]
    \label{eq:toy_model}
\end{equation}
This toy model in Equation \ref{eq:toy_model} does not aim to be physically precise, but a few helpful conclusions can be drawn from it. For a maser column with a mean length $s$, there are three sources of variability, with their corresponding time scales. The first is background variations $I_0(t)$ (hereafter background variability), the second is variations in the maser pump efficiency, measured by $\kappa_\nu$ (hereafter, pump variability), the third is the hydrodynamic component $\delta s(t)$ which changes the path length of the maser \citep[hereafter hydrodynamic variability,][]{1994ApJ...429..253G,2022ApJS..261...14L}. Theoretical work has studied maser time series in the case of rotation, line-of-sight overlap and shocks, which all fall under the umbrella of hydrodynamic variability \citep{2019MNRAS.486.4216G,2024IAUS..380..422G,2024MNRAS.530.3342G}. For 22 GHz water masers, positive or negative changes in $\kappa_\nu$ can be induced by changes in the external radiation field, kinetic temperature or density changes of the collisional partner H$_2$ \citep{2016MNRAS.456..374G,2022MNRAS.513.1354G}. If we take into account the fact that variations in $\delta s(t)$ can be faster than the separations between two observations at times $t_1$ and $t_2$, we cannot directly use the ratio of the two observations $I(t_1)/I(t_2)$ to draw any conclusions on the source of variability, due to the random component. But, if we have constraints on the time scales of $I_0(t)$, $t_{\rm background}$ or of $\kappa_\nu(t)$, $t_{\rm pump}$ and we know that either $t_{\rm background} \gg t_{\rm hydro}$ or $t_{\rm pump} \gg t_{\rm hydro}$, combined with multiple observations before and after the change in $I_0(t)$ or $\kappa_\nu(t)$, then the changes in the dispersion $\sigma_{\rm burst}/\sigma_{\rm pre}$ would probe changes in the physical conditions if $\langle\delta s(t)\rangle = 0$. Pump and hydrodynamic variability are not independent in terms of how they change $I_\nu$. According to Equation \ref{eq:toy_model}, pump variability can amplify or dampen the effects of hydrodynamic variability. We have reason to suspect that background variability was not the cause of the maser flare in NGC6334I (see Section \ref{subsubsec:change_background}), so we looked for changes in the dispersion before and after the accretion burst. 
\subsubsection{Testing Variance Increase in Water Maser Single-Dish Time Series} 
\label{subsub:testing_variance_change}
\begin{figure}[ht]
    \centering
    \includegraphics[width=0.435\textwidth]{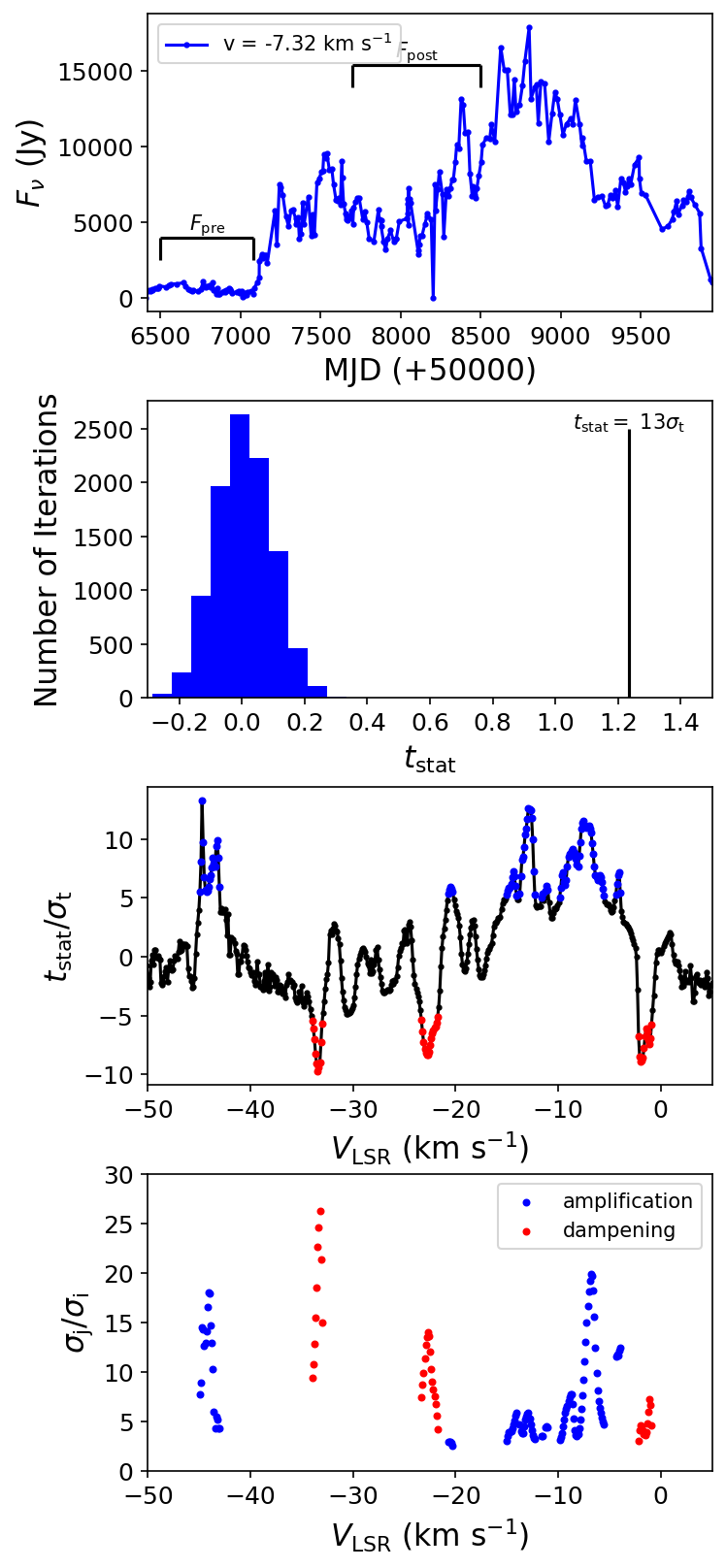}
    \caption{Permutation test of 22 GHz water maser long-term time series taken with HartRAO. Top panel: Example time series of the $-7.32$ km\,s$^{-1}$ velocity time series. The pre-burst and burst time windows are annotated. Second panel: Distribution of $t_{\rm stat}$ for $10^4$ iterations where the array is shuffled for the $-7.32$ km\,s$^{-1}$ time series. The value of the unshuffled time series is shown in the black line. Third panel: Value of the ratio of the unshuffled $t_{\rm stat}$ over the standard deviation of the shuffled distribution. The velocity channels which have statistically significant increased and decreased dispersion are shown in blue and red respectively. Bottom panel: Dispersion ratio for channels where there were statistically significant changes. The blue points is $\sigma_{\rm burst}/ \sigma_{\rm pre}$ where the dispersion increased while the red points are $\sigma_{\rm pre}/\sigma_{\rm burst}$ where the dispersion decreased.}
    \label{fig:variability_test}
\end{figure}
We used HartRAO monitoring data of 22 GHz water masers $F_{t_i, V_i} = F(t_i, V_i)$, where $F_\nu$ is the flux density in Jy, $t_i$ is the time in days and $V_i$ is the velocity channel. We segmented the data to pre-burst MJD $56500 < t_i <$ MJD $57080$ and burst MJD $57700 < t_i <$ MJD $58700$ data. We ignored the rise time at the onset of the accretion burst between MJD 57080 and MJD 57300. These data were binned into two arrays $\vec{F}_{\rm pre}$ and $\vec{F}_{\rm burst}$. The first panel of Figure \ref{fig:variability_test} shows the $-7.32$ km\,s$^{-1}$ HartRAO water maser time series, as well as the time windows used to construct $\vec{F}_{\rm pre}$ and $\vec{F}_{\rm burst}$ \citep[for details on the observations and data reduction, see ][]{2018MNRAS.478.1077M}. We applied a non-parametric permutation test outlined in \citet{tibshirani1993introduction} to see whether there was a statistically significant change in the dispersion before and after the accretion burst. The test assumes the data to be uncorrelated over time, and there to be no long-term trend influencing the data variance. In the case of only observing hydrodynamic variability in 22 GHz water masers, we can treat the data as uncorrelated as the path length changes in the masering gas are essentially random. To eliminate variance due to long-term trends, we take data in which the 6.7 GHz methanol masers do not change over time. We also fit a linear function of $\vec{F}_{\rm pre}$ and $\vec{F}_{\rm burst}$ vs time separately for each velocity channel to remove possible long-term (0.5 yr) linear trends. We stress that the residuals after the subtraction of the linear function are not instrumental noise, as we have SNR $ = 20 - 700$ in most cases, and the residuals are the intrinsic short-term variability in the water masers. We only are interested in short-term hydrodynamic variability, so removing the linear trends removes variance increase which might be due to pump variability with long timescales.  
The test statistic $t_{\rm stat}$ is given by:
\begin{equation}
    t_{\rm stat} = {\rm log}_{10}\Bigg(\frac{\sigma^2_{\rm post}}{\sigma^2_{\rm pre}}\Bigg)
    \label{eq:test_statistic}
\end{equation}
The value of $t_{\rm stat}$ will be positive if the variance increases between the pre-burst and post-burst cases, and negative if the variance decreases. If $\sigma^2_{\rm pre} = \sigma^2_{\rm post}$ then $t_{\rm stat} = 0$. 
The middle panel of Figure \ref{fig:variability_test} shows an example of this test for the $-7.32$ km\,s$^{-1}$ time series where there is a clear detection (13$\sigma_{\rm t}$) of increased dispersion. The bottom panel of Figure \ref{fig:variability_test} shows the results of this test for each velocity channel using the same time bins. This test does not suggest significant changes in variance for most of the velocity channels, which gives some confidence to its robustness. The bottom panel of Figure \ref{fig:variability_test} shows the ratio of amplification or dampening between the pre-burst and post-burst periods. We report individual velocity channels as there can be multiple spatially independent maser features in a single velocity feature of a single-dish observation. The velocity channels at $-44$ km\,s$^{-1}$, $-7$ km\,s$^{-1}$ and $-4$ km\,s$^{-1}$ showed an order of magnitude increase in $\sigma_{\rm burst}/\sigma_{\rm pre}$. On the other hand velocity channels at $-33$ km\,s$^{-1}$, $-22$ km\,s$^{-1}$ and $-1$ km\,s$^{-1}$ showed dampening of a factor 26, 15 and 7 in their dispersions respectively. Using current and previous interferometric maps of 22 GHz water masers, we identified the dominant features contributing to $\sigma_{\rm burst}/\sigma_{\rm pre}$ changes in each velocity range. Figure \ref{fig:velocities_per_association} and Table \ref{tab:maser_features} are helpful for this comparison. The increase in $\sigma_{\rm burst}/\sigma_{\rm pre}$ for the $-7$ km\,s$^{-1}$ and $-4$ km\,s$^{-1}$ can be attributed to the flaring of CM2-W2, which is the dominant maser feature in the field after 2015. The increase in the variance for the $-44$ km\,s$^{-1}$ might be associated with the high-velocity features in CM2-W2 after 2015.5. On the other hand, the decrease in the variance of the features around $-33$ km\,s$^{-1}$, $-22$ km\,s$^{-1}$ and $-1$ km\,s$^{-1}$ can be attributed to the dampening of features in UCHII-W3, UCHII-W1 and MM1-W1 respectively (Figure \ref{fig:velocities_per_association}). These comparisons show that variance changes before and during an accretion burst in 22 GHz water maser monitoring data can be a useful probe of the excitation and destruction of water maser features. Testing on theoretically modelled time series, and long-term monitoring data for other sources help improve our understanding of the strengths and weaknesses of this analysis.
These results support our earlier classification of water maser variability types. Pump efficiency variations can amplify or dampen random flux density fluctuations caused by chaotic fluid motion in shocked regions outlined by \citet{1994ApJ...429..241G,1994ApJ...429..253G}. This occurs through changes in maser depth per path length. This analysis and the VLBI observations reported in this work imply that the variable radiation fields due to the accretion burst caused pump or background variability (dampening and amplification) in the water masers in NGC6334I. In the case of a water maser flare, chance line of sight overlap can be distinguished from pump variability by using the method in this section to check whether or not there are changes in the dispersion of the maser time series. Although the dispersion test possibly indicates pump variability, it does not explain the cause of the variability. We will consider possible sources of non-hydrodynamic variability in Section \ref{subsubsec:radiative_heating_h2}.

These results also show that single water maser observations in a time series cannot be predicted, but that their statistics hold information on the underlying physics. Time-dependent background radiation and changes in the maser depth are important components, but the stochastic hydrodynamic variability cannot be ignored. The distribution of time scales for hydrodynamic variability would need to be determined from simulations of the complex fluid flow in the shock fronts in which these masers appear. Further, it should be taken into consideration that single-dish time series rarely only observe the radiation from a single maser feature and that single maser features have been seen to drift in velocity. Characterizing these effects could enhance the analysis of long-term ($>$10 year) single-dish water maser monitoring.
\subsection{How did the Accretion Burst Cause the Water Masers to Flare in CM2-W2?}
\label{subsec:water_maser_flare_cm2w2}
How does the accretion burst lead to a positive correlation between the radiation field strength and flux density of water masers in CM2-W2? Here we consider possible explanations: Amplification of background radiation, radiatively pumped 22 GHz water masers, or radiative heating of the collisional partner.
\subsubsection{Change in Background Continuum}
\label{subsubsec:change_background}
The change in the background continuum can be tested with pre-burst and burst centimetre continuum measurements of CM2-W2. The pre-burst 22 GHz (1.5 cm) continuum was not detected at the $3\sigma = 0.30$ mJy beam$^{-1}$ level with JVLA \citep{2016ApJ...832..187B}, and the post-burst 1.5 cm continuum was not reported. On the other hand, for synchrotron continuum emission, the flux density at 5 cm and 1.5 cm are proportional. The 5 cm continuum was detected, and it decreased from 0.36 mJy beam$^{-1}$ in 2011 to 0.19 mJy beam$^{-1}$ in 2016.9 \citep{2016ApJ...832..187B,2018ApJ...854..170H}. It does not seem that an increase in the background continuum is responsible for the maser flare in CM2-W2.
\subsubsection{Radiatively Pumped Water Masers}
It has been proposed that strong infrared radiation could radiatively pump bright 22 GHz water masers  \citep{2022MNRAS.513.1354G}. In this scheme, although water masers are dampened by increased dust temperature in most densities and kinetic temperatures, there is a part of the parameter space, $10^{4} <$ n $_{{\rm H}_2{\rm O}}$ $< 10^6$ cm$^{-3}$, T$_{\rm kin} < 500$ K and T$_{\rm dust} > 900$ K, where maser optical depth scales very quickly with T$_{\rm dust}$. Radiatively pumped 22 GHz water masers then correspond to "cold-gas hot-dust" regions.
We consider this proposal of radiatively pumped water masers to be insufficient in our case for two reasons. First, CM2-W2 is at the edge of a protostellar jet, which has high T$_{\rm kin}$ compared to T$_{\rm dust}$, as seen in maser morphology, proper motion direction and thermal line emission \citep{2018ApJ...866...87B,2021ApJ...908..175C,2021ApJ...912L..17H}. Second, this proposal is valid if there was a significant change in the dust temperature from the pre-burst to burst epochs, but the changes in millimetre dust emission were not observed in CM2-W2 \citep{2017ApJ...837L..29H}.
\subsubsection{Radiative Heating of H$_2$}
\label{subsubsec:radiative_heating_h2}
We also consider the possibility that the water masers stayed collisionally pumped, but that the accretion burst increased the maser pump efficiency ($\kappa_\nu$ in Equation \ref{eq:toy_model}) enough to cause the maser flare. Analysis of the effect of the accretion burst on masers can be broken up into three parts. The first is the spectrum of the bursting YSO and its radiative transfer to the masering gas. Followed by the changes in the physical conditions of the masering gas, such as heating and chemical effects. Lastly, one can consider the maser properties in the irradiated gas. We suggest what might be the most reasonable explanation given the observations of NGC6334I-MM1 and water maser models. Our analysis here is relatively simple, and we leave more detailed calculations on each of the steps above for future work.
 
The accretion burst led to a bolometric luminosity increase by a factor of $16\pm 4.4$ in MM1 \citep{2021ApJ...912L..17H}. All continuum observations are in the far-infrared to submillimetre wavelengths which only observed reprocessed dust emission \citep{2017ApJ...837L..29H,2018ApJ...866...87B,2021ApJ...912L..17H}. We assume that the radiation that terminates at CM2-W2 is weakly attenuated by dust and preserves its original spectral shape (Section \ref{subsub:discussion_source_geometry}). The main spectral components are the photosphere \citep[mainly UV for T$_{\rm eff} = 16000 $K,][]{2021ApJ...912L..17H} and the accretion disk \citep[infrared and non-thermal X-rays, e.g.][]{2019A&A...630A..84D}. We rule out infrared radiation as an explanation as it would rather dim the water masers in CM2-W2 as in NGC6334I-MM1 and G107.298$+$5.639 \citep{2016MNRAS.459L..56S,2018ApJ...866...87B}.
 
High-energy radiation would affect the chemistry and ionization state of the masering gas. Maser models for C-shocks and J-shocks neglect external ultraviolet radiation, so it is unclear what higher-order effects this radiation would have on the masers \citep{1996ApJ...456..250K,2013ApJ...773...70H}. We can see how the radiation might affect T$_{\rm kin}$ of the masering gas. We use simple energy arguments to make an order of magnitude estimate for $\Delta {\rm T}_{\rm kin}$ due to the accretion burst. The energy balance equation:
\begin{equation}
    \frac{dU}{dt} = \Gamma - \Lambda
    \label{eq:energy_balance}
\end{equation}
where $U$ is the internal energy of the gas in ergs while $\Gamma$ and $\Lambda$ are the heating and cooling rates. For simplicity, we will assume $\Gamma \gg \Lambda$. In C-shocks with water masers, the main source of cooling is water maser emission, the dust and H$_2$ thermal line cooling being weaker by an order of magnitude \citep{1996ApJ...456..250K}. We can integrate Equation (\ref{eq:energy_balance}) to get:
\begin{equation}
    U = U_0 + \int_{t_{\rm pre}}^{t_{\rm post}}\Gamma(t) dt    \label{eq:internal_energy_time_dependence}
\end{equation}
Where $U_0$ is the energy of the masering gas before the burst, $t_{\rm pre}$ the time before the burst, $t_{\rm post}$ the time of the peak 6.7 GHz methanol and $t_{\rm rise} = t_{\rm post} - t_{\rm pre}$. If we assume that the masering gas has an optical depth so that some constant fraction $\eta$ of the incident radiation is absorbed as heat, then $\Gamma = \eta F dS$. The incident $F$ is from the YSO at MM1B, and $dS$ is the area of the maser exposed to $F$. The brightest masers are perpendicular to the shock propagation direction \citep{2013ApJ...773...70H}. For a cylindrical maser perpendicular to the YSO, $dS  = a r^2_{\rm maser}$ where $r_{\rm maser}$ is the maser radius \citep[$\sim 1$ au,][]{1994ApJ...429..253G} and $a$ is the maser aspect ratio \citep[$10 - 100$,][]{2013ApJ...773...70H}. 

The YSO flux is given by $F = L_*/4\pi R^2$, with $L_*$ the accretion and photospheric bolometric luminosity of the YSO, and $R$ is the distance from the YSO and the maser (2800 au). If we assume an ideal equation of state $U = mC_{\rm V}{\rm T}_{\rm kin}$, where $m = {\rm n}_{{\rm H}_2}V = \mu_{ {\rm H}_2} {\rm n}_{{\rm H}_2}ar^3_{\rm maser}$ is the mass of the maser. With $\mu_{{\rm H}_2}$ the mass of a hydrogen molecule and n$_{{\rm H}_2} \sim 10^{8-10}$ cm$^{-3}$ the hydrogen number density. The heat capacity per unit mass at constant volume, $C_{\rm V}$, can be obtained from the heat capacity per mole at constant pressure ($C_{\rm P}$) from \citet{chase1985janaf} with $\gamma = 7/5$ for an ideal diatomic gas and $T$ the kinetic temperature, Equation (\ref{eq:internal_energy_time_dependence}) becomes:
\begin{equation}
    \Delta {\rm T}_{\rm kin} = \frac{\eta\gamma N_{\rm A}}{4\pi R^2 \cdot {\rm n}_{{\rm H}_2} r_{\rm maser} C_{\rm P}}\int^{t_{\rm post}}_{t_{\rm pre}} L_*(t) dt
    \label{eq:temperature_time_dependence}
\end{equation}
We can assume that $L_*(t)$ has a time dependence similar to the $-7.26$ km\,s$^{-1}$ 6.7 GHz methanol maser time series \citep{2018MNRAS.478.1077M}. The pre-burst and burst luminosity has been estimated with $L_*(t_{\rm pre}) = 4300\pm900L_\odot$ and $L_*(t_{\rm post}) = 49000\pm8000 L_\odot$. For a step function in luminosity, the integral in Equation \ref{eq:temperature_time_dependence} becomes $t_{\rm rise}\Delta L_*$. Keeping n$_{{\rm H}_2}$ and $\eta$ as free parameters Equation \ref{eq:temperature_time_dependence} becomes:
\begin{equation}
    \Delta {\rm T}_{\rm kin} =  2.8\times 10^5\frac{\eta}{{\rm n}_8} {\rm K}
\end{equation}
where n$_8$ is n$_{{\rm H}_2}$ in units of $10^8$ cm$^{-3}$. This shows that even if the (non-maser) cooling $\Lambda$ is half of the heating $\Lambda = 0.5 \Gamma$, and the heating efficiency $\eta$ is less than one percent $\eta < 0.01$,  the gas in CM2-W2 would heat by 10$^3$ K.

To estimate $\eta$, the spectral shape of the incident radiation is essential as H$_2$ has wavelength-dependent opacity. Modelling of interstellar gas heated with X-rays suggests that 1 keV X-rays can have heating efficiencies of up to 10$\%$ \citep{2012ApJ...756..157G}. On the other hand H$_2$ has a non-negligible opacity to ultraviolet at the densities (n$_{{\rm H}_2} \sim 10^{8-10}$ cm$^{-3}$) in which water masers are found \citep{1989ApJ...338..197S}. The spectrum of an accretion-bursting HMYSO is still an open question. However, we consider it possible that a bursting HMYSO could radiate enough high-energy radiation to produce the heating efficiency required for a significant increase in T$_{\rm kin}$.

There is empirical reason to think that CM2-W2 is being heated up more than the other regions, and that is the detection of 321 GHz water masers at the south of the bow-shock (Figure \ref{fig:321GHz_masers}). These masers require high gas temperatures T$_{\rm kin} \sim 1000$ K.

After we consider the change in T$_{\rm kin}$, we need to consider the response of the water maser. There are differences in the model predictions to the maser response to gas heating. The J-shock model of \citet{2013ApJ...773...70H} has a weak dependence of maser brightness temperature to gas kinetic temperature. In this model, to have a maser flare in a single feature with a factor of four, the gas temperature needs to increase by $\sim 300$ K. On the other hand, in the model of \citet{2016MNRAS.456..374G} the maser depth responds differently to temperature changes at different densities. For hydrogen number densities 3 $\times 10^{8}$ cm$^{-3}$, the maser depth is very weakly dependent on T$_{\rm kin}$, while at very high densities between 3 $\times 10^{8}$ cm$^{-3}$ and 3 $\times 10^{10}$ cm$^{-3}$, maser depth changes rapidly between T$_{\rm kin} = 200$ K and T$_{\rm kin} = 500$ K. This can be most clearly seen in Figure 1 of \citet{2022MNRAS.513.1354G}, where for T$_{\rm dust} =50$ K, maser depth changes by a factor of two between T$_{\rm kin} = $ 250 K and 350 K. As maser depth goes at least linearly (for saturated masers) or exponentially (for unsaturated masers), small temperature changes between 200 K and 400 K at hydrogen number densities between 10$^8$ cm$^{-3}$ and 10$^{10}$ cm$^{-3}$ can adequately explain the maser flare in the unsaturated case. The calculations of \citet{1997MNRAS.285..303Y} (independent of shock type) indicate that water masers have a sharp rise in maser depth between 500 and 1000 K, but that additional increase in T$_{\rm kin}$ after 1000 K would not help the maser action.

These model calculations show that under negligible non-maser cooling and relatively low heating efficiencies, the accretion burst can cause temperature jumps that would amplify the maser emission in CM2-W2 to observed levels. Our explanation for water maser flares in accretion-bursting sources is similar to that of \citet{2024MNRAS.530.3342G}, although their argument follows different details and they focus on the burst in G358.93-0.03-MM1.

\section{Conclusion} 
\label{sec:conclusion}
We did multi-epoch VLBI observations of 22 GHz water masers in 2014$-$2016 with VERA. These observations cover epochs before, during and after the onset of the accretion burst in NGC6334I. We also report the first detection and imaging of 321 GHz water masers in this source with ALMA. These multi-epoch observations allow us to track maser evolution throughout the accretion burst, providing unique insights into how the burst affects different maser environments within the complex source structure.
\begin{enumerate}
    \item In CM2-W2 associated with a bow shock at the northern edge of the NW-SE jet, we found large-scale ($> 50$ au) and small-scale ($\sim $1 au) variations in the water masers. The large-scale size increase of the maser region and the velocity-dependent temporal variations suggest a propagating effect from the accretion burst, likely due to increased energy input into the surrounding medium. This expansion and the detection of new maser features at different velocities over time provide direct evidence of the burst's impact on the maser environment.
    \item The masers in MM1-W1 close to the bursting source and UCHII-W1$-$3 at the southern edge of two protostellar jets show complex variability patterns that reflect the intricate source structure and different shock types. MM1-W1 exhibited negative pump variability consistent with high dust temperatures. UCHII-W3, despite being radiatively connected to MM1B, did not show similar flaring to CM2-W2, possibly due to differences in shock types (J-shocks vs. C-shocks). UCHII-W2 showed gradual dimming, while UCHII-W1 displayed complex behaviour. These varied water maser responses highlight the importance of source geometry, shock types, and the different effects of thermal, infrared, and high-energy components from the accretion burst. 
    \item We estimated absolute water maser proper motions in the pre-burst ($2014.72 -2015.28$) and burst ($2015.88 - 2016.19$) cases. Low-velocity $v < 40$ km\,s$^{-1}$ proper motions were detected in CM2-W2 and MM1-W1, likely from non-dissociative C-shocks. Alternatively, the proper motions in UCHII-W1, UCHII-W2 and UCHII-W3 were consistent with a high-velocity $v > 70$ km\,s$^{-1}$ dissociative J-shocks. No significant changes were found in the mean velocity of the proper motions. The consistency in mean proper motion velocities before and during the burst provides further evidence that observed changes in maser distribution are primarily due to excitation effects rather than physical motions. The presence of excitation effects underscores the importance of careful interpretation of maser proper motions, especially in dynamic environments experiencing accretion bursts.
    \item We used a toy maser model to introduce the concepts of background, hydrodynamic and pump variability. Our VLBI observations of individual maser features complement this analysis by providing high-resolution spatial information, allowing us to connect changes in maser variance to specific regions within NGC6334I. With long-term single-dish monitoring observations from HartRAO, we identified statistically significant changes in the variance of the time-dependent flux density in certain velocity channels. Variance changes after the burst are associated with changes in the maser pump efficiency, which amplifies hydrodynamic variability. We propose the changes in the variance of a collisionally pumped maser flux density time series as a way to identify pump variability in water masers.
    \item We considered possible explanations for the contemporaneous flaring of collisionally pumped 22 GHz water masers and 6.7 GHz methanol masers in NGC6334I. The spatial distribution and intensity changes of masers observed in our VLBI monitoring support the hypothesis of radiative heating from high-energy radiation propagating through low-optical depth cavities. These observations highlight that the maser response of irradiated shocks deserve further theoretical investigation.
\end{enumerate}
In conclusion, our multi-epoch VLBI observations, combined with theoretical analysis, reveal the complex interplay between accretion bursts, source structure, and water maser behaviour in NGC6334I. Future work in the theoretical investigation of masers in irradiated shocks would be useful for the interpretation of water masers in future accretion bursts. Additionally, the contemporaneous flaring of collisionally pumped and radiatively pumped transitions in bursting sources prompts observational investigation in infrared and high-energy regimes.
\begin{acknowledgements}
      We thank the anonymous referee for thoughtful comments which improved the quality and clarity of this manuscript. JMV acknowledges funding from the National Research Foundation (UID: 134192). JMV thanks Alexander Rawlings for valuable input on statistical methods. JMV and MJ acknowledge support from the Research Council of Finland grant No. 348342. T.H. is financially supported by the MEXT/JSPS KAKENHI grant Nos. 17K05398, 18H05222, and 20H05845. E.I.V. and A.M.S. acknowledge support from the Russian Science Foundation, project No. 23-12-00258. JOC acknowledges financial support from the South African Department of Science and Innovation's National Research Foundation under the ISARP RADIOMAP Joint Research Scheme (DSI-NRF Grant Number 150551). This paper makes use of the following ALMA data: ADS/JAO.ALMA$\#$2018.1.00024. ALMA is a partnership of ESO (representing its member states), NSF (USA) and NINS (Japan), together with NRC (Canada), NSTC and ASIAA (Taiwan), and KASI (Republic of Korea), in cooperation with the Republic of Chile. The Joint ALMA Observatory is operated by ESO, AUI/NRAO and NAOJ. HartRAO forms part of SARAO, which is a National facility operating under the auspices of the
National Research Foundation (NRF), South Africa.
\end{acknowledgements}
\appendix
\section{Proper Motion Considerations Unrelated to the Accretion Burst}
\label{subsec:discussion_pm}
\subsection{Comparison with \citet{2021ApJ...908..175C}}
The proper motions that we estimated show a characteristic bipolar outflow pattern. Relative proper motions for this source were also estimated by \citet{2021ApJ...908..175C} with KaVA. Their observations had a higher angular and spectral resolution and better sensitivity but only covered three epochs a year after the 2015.0 onset of the accretion burst (spanning $2015.89 - 2016.01$). On the other hand, our seven observations span times from before the burst (2014.7) until a year after the burst (2016.2). The largest systematic difference between our results and those of \citet{2021ApJ...908..175C} is that they did not shift their proper motions to an absolute velocity reference frame. They reported a large ($\sim 120$ km\,s$^{-1}$) northward transverse speed for CM2-W2 and northward pointing proper motions in UCHII-W1 contrary to our observations and \citet{2018ApJ...866...87B}. This led them to suspect that the MM3-UCHII might be affecting the proper motions of masers in UCHII-W2 and UCHII-W3. This is due to their proper motions being in the reference frame of a bright maser in UCHII-W1, and therefore their relative proper motions should be shifted to an absolute velocity frame to be physically interpreted. They also calculated proper motions for spots (individual velocity channels), rather than maser features (see Section \ref{subsubsec:spots_and_features}). Their Velocity-Variance Covariance Matrix \citep[VVCM][]{1993ApJ...406L..75B,2000ApJ...533..893B} estimate of the outflow and inclination angle of the NW-SE outflow is independent of velocity frame, and is therefore robust. Our high-velocity ($\sim100$ km\,s$^{-1}$) proper motions for UCHII-W1 are consistent with the bulk motion observed by \citet{2018ApJ...866...87B} between 2011 and 2017.8.
\subsection{Reliability of Proper Motions}
We do not expect the underlying gas motion to be affected by the accretion burst (see the timing argument at the end of Section \ref{subsub:cm2_w2_variability}). Changes in the proper motions should rather be due to changes in pumping conditions, rather than real gas motion. This is analogous but with a different physical mechanism than the propagation of 6.7 GHz methanol masers at high velocities seen by \citet{2020NatAs...4..506B}. The water masers' proper motions in CM2-W2, MM1-W1 and UCHII-W1 show more variation in their vector direction as those in some other sources \citep[e.g. ][]{2016MNRAS.460..283B}. It has been argued by \citet{2006A&A...447L...9G} that water maser proper motions reliably trace gas motion. Their argument is empirical, and based on the ordered motion of water masers in  G24.78$+$0.08 over three epochs. Their argument only shows that some (one source) water masers trace gas motion, but not all. The empirical (see Section \ref{sec:results}) and theoretical (see Section \ref{subsec:variability_mechanisms}) considerations in this work indicate that great care should be taken when interpreting water maser proper motions as physical gas motion. For water maser proper motions to be reliable observable, robust inter-epoch identification methods for maser features should be considered. It should be considered whether the three types of variability (background, pump and hydrodynamic, defined in Section \ref{subsec:variability_mechanisms}) are on longer or shorter time-scales than the observations. 
\section{Supporting Figures}
\begin{figure*}
    \centering
    \includegraphics[width=\textwidth]{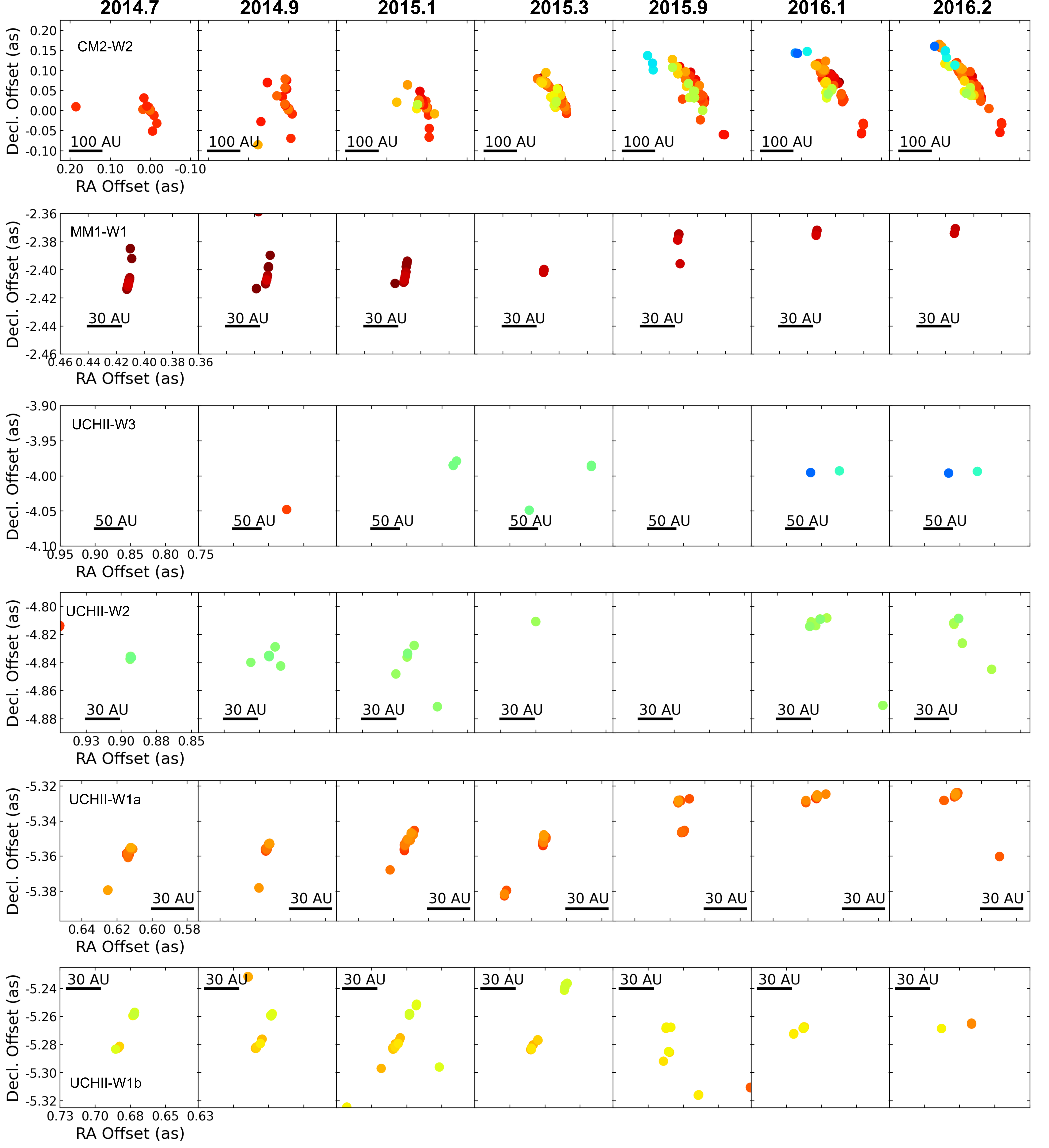}
    \caption{Water maser spot positions epochs 2014.7$-$2016.2. Each horizontal panel shows a different association over time, with the name of the region in the leftmost panel. The spatial scale is constant over epochs for each association. The dates of observation are shown above the top panel. A linear scale is shown in each panel. The colour scale is the same as Figure \ref{fig:vera_spotmaps}.}
    \label{fig:vera_subregions}
\end{figure*}

%
\bibliographystyle{aa} 
\bibliography{bibliography} 
%

\end{document}